\documentclass[a4paper,journal,twoside]{IEEEtran}

\ifCLASSINFOpdf
\else
\fi

\usepackage[cmex10]{amsmath}
\usepackage{amssymb}
\usepackage{mathtools}
\usepackage{epsfig}
\usepackage{epstopdf}
\usepackage{subfigure}
\usepackage{booktabs}
\usepackage{multirow}
\usepackage{blindtext}
\usepackage{algorithm}
\usepackage{algorithmic}
\usepackage{color}
\usepackage{amsmath}
\usepackage{graphicx}
\usepackage{standalone}

\usepackage[noadjust]{cite}
\newsavebox{\ieeealgbox}

\usepackage{array}

\usepackage{mdwmath}
\usepackage{mdwtab}

\usepackage{stfloats}
\fnbelowfloat

\DeclareMathOperator{\RSNR}{SNR_{rec}}

\newcommand{\ACal}{\mathcal{A}}
\newcommand{\Ab}{\mathbf{A}}
\newcommand{\Bb}{\mathbf{B}}
\newcommand{\Cb}{\mathbf{C}}
\newcommand{\Ib}{\mathbf{I}}
\newcommand{\Pb}{\mathbf{P}}
\newcommand{\Qb}{\mathbf{Q}}
\newcommand{\Mb}{\mathbf{M}}
\newcommand{\Sb}{\mathbf{S}}
\newcommand{\Ub}{\mathbf{U}}
\newcommand{\Vb}{\mathbf{V}}
\newcommand{\bb}{\mathbf{b}}
\newcommand{\Wb}{\mathbf{W}}
\newcommand{\Xb}{\mathbf{X}}
\newcommand{\xb}{\mathbf{x}}

\newcommand{\yb}{\mathbf{y}}
\newcommand{\zb}{\mathbf{z}}
\newcommand{\Yb}{\mathbf{Y}}
\newcommand{\Zb}{\mathbf{Z}}
\newcommand{\Zerb}{\mathbf{0}}
\newcommand{\Xbh}{\widehat{\Xb}}
\newcommand{\Ybh}{\widehat{\Yb}}
\newcommand{\Zbh}{\widehat{\Zb}}
\newcommand{\Xbd}{\Xb_{\delta}}

\newcommand{\Rbb}{\mathbb{R}}
\newcommand{\DimDef}{\Rbb^{n_1 \times n_2}}
\newcommand{\NCal}{\mathcal{N}}

\newcommand{\Sbb}{\mathbb{S}}
\newcommand{\Sbbn}{\Sbb^n}
\newcommand{\Sbbp}{\Sbb_{+}}
\newcommand{\LOp}{\ACal: \DimDef \rightarrow \Rbb^m}
\newcommand{\nDef}{n=\min(n_1,n_2)}

\newcommand{\UDelmOne}{\lceil \Delta-1 \rceil}

\newcommand{\FCal}{\mathcal{F}}
\newcommand{\SigMat}{\boldsymbol{\Sigma}}

\newcommand{\LamVec}{\boldsymbol{\lambda}}
\newcommand{\LamVecD}{\LamVec^{\downarrow}}
\newcommand{\LamVecA}{\LamVec^{\uparrow}}
\newcommand{\Sigb}{\boldsymbol{\sigma}}

\newcommand{\LambXb}{\boldsymbol{\lambda}(\Xb)}
\newcommand{\LambYb}{\boldsymbol{\lambda}(\Yb)}
\newcommand{\SigbXb}{\Sigb(\Xb)}

\newcommand{\fdel}{f_{\delta}}
\newcommand{\Fdel}{F_{\delta}}
\newcommand{\Hdel}{H_{\delta}}
\newcommand{\sumn}{\sum_{i=1}^{n}}
\newcommand{\NullDef}{\mathcal{N}(\ACal)}
\newcommand{\NullDefmZ}{\NullDef\setminus \{\Zerb\}}
\newcommand{\SNR}{\text{SNR}_{\text{rec}}}

\DeclareMathOperator{\rank}{rank}

\DeclareMathOperator{\diag}{diag}
\DeclareMathOperator{\vect}{vec}

\DeclareMathOperator*{\argmin}{argmin}

\DeclareMathOperator{\trace}{trace}

\begin{document}
\title{Iterative Concave Rank Approximation for Recovering Low-Rank Matrices}
%
%
%

\author{Mohammadreza~Malek-Mohammadi,~\IEEEmembership{Student Member,~IEEE}, Massoud~Babaie-Zadeh,~\IEEEmembership{Senior~Member,~IEEE}, and~Mikael~Skoglund,~\IEEEmembership{Senior~Member,~IEEE}
\thanks{This work was supported in part by the Iran Telecommunication Research Center (ITRC) under contract No.~500/11307 and the Iran National Science Foundation under contract No.~91004600. The work of the first author was supported in part by the Swedish Research Council under contract 621-2011-5847 and a travel scholarship from Ericsson Research during his visit at KTH.}
\thanks{M. Malek-Mohammadi and M. Babaie-Zadeh are with the Electrical Engineering Department,
Sharif University of Technology, Tehran 1458889694, Iran (e-mail: m.rezamm@ieee.org; mbzadeh@yahoo.com).}
\thanks{M. Skoglund is with the Communication Theory Lab, KTH, Royal Institute of Technology, Stockholm, 10044, Sweden (e-mail: skoglund@ee.kth.se).}}

\maketitle

\begin{abstract}
In this paper, we propose a new algorithm for recovery of low-rank matrices from compressed linear measurements. The underlying idea of this algorithm is to closely approximate the rank function with a smooth function of singular values, and then minimize the resulting approximation subject to the linear constraints. The accuracy of the approximation is controlled via a scaling parameter $\delta$, where a smaller $\delta$ corresponds to a more accurate fitting. The consequent optimization problem for any finite $\delta$ is nonconvex. Therefore, in order to decrease the risk of ending up in local minima, a series of optimizations is performed, starting with optimizing a rough approximation (a large $\delta$) and followed by successively optimizing finer approximations of the rank with smaller $\delta$'s.
To solve the optimization problem for any $\delta > 0$, it is converted to a new program in which the cost is a function of two auxiliary positive semidefinete variables. The paper shows that this new program is concave and applies a majorize-minimize technique to solve it which, in turn, leads to a few convex optimization iterations. This optimization scheme is also equivalent to a reweighted Nuclear Norm Minimization (NNM), where weighting update depends on the used approximating function. For any $\delta > 0$, we derive a necessary and sufficient condition for the exact recovery which are weaker than those corresponding to NNM. On the numerical side, the proposed algorithm is compared to NNM and a reweighted NNM in solving affine rank minimization and matrix completion problems showing its considerable and consistent superiority in terms of success rate, especially, when the number of measurements decreases toward the lower-bound for the unique representation.
\end{abstract}

\begin{IEEEkeywords}
Affine Rank Minimization (ARM), Matrix Completion (MC), Nuclear Norm Minimization (NNM), Rank Approximation, Null-Space Property (NSP).
\end{IEEEkeywords}

%
\IEEEpeerreviewmaketitle

\section{Introduction}
\IEEEPARstart{R}{ecovery} of low-rank matrices from underdetermined linear measurements, generalization of the recovery of sparse vectors from incomplete measurements, has become a topic of high interest within the past few years in signal processing, control theory, and mathematics. This problem has many applications in various areas of engineering. For example, collaborative filtering \cite{CandP10}, ultrasonic tomography \cite{ParhKOV13}, direction-of-arrival estimation \cite{MaleJOKB14}, and machine learning \cite{AmitFSU07} are some of these applications. For more comprehensive lists of applications, we refer the reader to \cite{RechFP10,CandR09,CandP10}.

Mathematically speaking, the rank minimization (RM) problem under affine equality constraints (linear measurements), which we refer to as ARM, is described by
\begin{equation}  \label{RM}
\min_{\Xb}\rank(\Xb) ~ \text{ subject to } ~ \ACal(\Xb)=\bb,
\end{equation}
in which $\Xb \in \DimDef$ is the optimization variable, $\LOp$ is a linear measurement operator, and $\bb \in \Rbb^m$ is the vector of available measurements. The constraints are underdetermined meaning that $m < n_1 n_2$ or more often $m \ll n_1 n_2$. The above formulation has the so-called matrix completion (MC) problem as an important instant corresponding to
\begin{equation} \label{MC}
\min_{\Xb} \rank(\Xb) ~ \text{ subject to } ~ [\Xb]_{ij} = [\Mb]_{ij},~~\forall(i,j) \in \Omega,
\end{equation}
where $\Mb \in \DimDef$ is the matrix whose elements are partially known, $\Omega \subset \{1,2, ..., n_1\} \times \{1,2, ..., n_2\}$ is the set of the indexes of known entries of $\Mb$, and $[\Xb]_{ij}$ designates the $(i,j)$th entry of $\Xb$. When $\rank(\Xb^*)$ is sufficiently low and $\ACal$ has some favorable properties, $\Xb^*$ is a unique solution to \eqref{RM} \cite{RechFP10,OymaH10}.

Nevertheless, \eqref{RM} is in general NP-hard and very challenging to solve \cite{ChistG84}. A well-known replacement is nuclear norm minimization (NNM) approach \cite{RechFP10} formulated as
\begin{equation}  \label{NNM}
\min_{\Xb} \|\Xb\|_* ~ \text{ subject to } ~ \ACal(\Xb)=\bb,
\end{equation}
where $\|\Xb\|_*$ denotes the nuclear norm of $\Xb$ equal to the sum of singular values of $\Xb$. It has been shown that, under more restrictive assumptions on the rank of $\Xb^*$ or properties of $\ACal$, \eqref{RM} and \eqref{NNM} share the same unique solution $\Xb^*$ \cite{RechFP10}.

When measurements are contaminated by additive noise, one way to robustly find a solution, is to update \eqref{RM} to
\begin{equation} \label{RMNoisy}
\min_{\Xb}\rank(\Xb)~ \text{ subject to } ~ \| \ACal(\Xb)-\bb \|_2 \leq \epsilon,
\end{equation}
where $\|\cdot\|_2$ denotes the $\ell_2$ norm and $\epsilon$ is some constant not less than noise power. Accordingly, \eqref{NNM} is also converted to
\begin{equation}  \label{NNMNoisy}
\min_{\Xb} \|\Xb\|_* ~ \text{ subject to } ~ \| \ACal(\Xb)-\bb \|_2 \leq \epsilon.
\end{equation}
Again, under some mild conditions on $\rank(\Xb^*)$ and properties of $\ACal$, the solution of \eqref{NNMNoisy} is close to the solution of \eqref{RMNoisy} in terms of their distance measured by the Frobenius norm \cite{CandP11}.

There are some other approaches to solve the ARM problem. Some of them are efficient implementations of NNM such as FPCA \cite{MaGC11}, APG \cite{TohY10}, and SVT \cite{CaiCS10}. Some others are based on generalization of the methods already proposed for sparse recovery in the framework of compressive sampling (CS) \cite{Bara07} like ADMiRA \cite{LeeB10} and SRF \cite{MaleBAJ14} which extend CoSaMP \cite{NeedT09} and SL0 \cite{MohiBJ09} to the matrix case, respectively.

Despite the convexity of the NNM program, there is a large gap between the sufficient conditions for the exact and robust recovery of low-rank matrices using \eqref{RM} and \eqref{NNM} \cite{MohaFH11}. To narrow this gap, we introduce a novel algorithm based on successive and iterative minimization of a series of nonconvex replacements for \eqref{RM}. Although our theoretical analysis shows that global minimization of each replacement in the series recovers solutions at least as good as NNM approach does, our numerical simulations demonstrate that the proposed chain of minimizations results in considerable reduction in the number of samples required to recover low-rank matrices. This improvement is achieved at the cost of higher computational complexity. Nevertheless, in some applications of MC and ARM, like magnetic resonance imaging \cite{LingHDJ11,MajuW12}, quantum state tomography \cite{GrosLFBE10}, and system identification and low-order realization of linear systems \cite{RechFP10}, reduction in the number of samples can be very beneficial, whereas complexity is not a big concern.

We improve over the method of SRF in \cite{MaleBAJ14,GhasMBJ11} which uses a class of nonconvex functions to approximate the rank function and iteratively minimizes the resulting approximation. In \cite{MaleBAJ14}, the nonconvex cost function scales with a parameter $\delta$ which reflects the accuracy. The smaller $\delta$, the more accurate approximation of the rank. SRF starts with a large $\delta$ and decreases it gradually to gain more accurate approximations of \eqref{RM} and successively optimizes the series of approximations. Numerical simulations show superiority of SRF to NNM and some other sate-of-the-art algorithms in both MC and ARM problems\cite{MaleBAJ14}; however, since the collection of exploited functions lack the subadditivity property, there is no guarantee that globally minimizing the proposed replacement of \eqref{RM} for any $\delta > 0$ leads to the exact recovery of the minimum-rank solution except for the asymptotic case of $\delta \to 0$.

In this paper, we use a class of subadditive approximating functions instead. As a result, a necessary and sufficient condition for the exact recovery is derived for any $\delta > 0$ which is weaker than that of NNM. In addition, we show that, under the same conditions, all matrices of rank equal or higher than what is guaranteed by \eqref{NNM} can be uniquely recovered by globally minimizing the cost function for any nonzero $\delta$. Another interesting result shows that as $\delta \to \infty$, the proposed optimization coincides with NNM.

To solve the resulting optimization problems, similar to \cite{FazeHB03}, we convert them to other programs in which the domain of the approximating functions is limited to the cone of Positive SemiDefinite (PSD) matrices. In this fashion, the rank approximating functions are concave and differentiable, so we use a Majorize-Minimize (MM) technique consisted of a few SemiDefinite Programs (SDP) to optimize them. Hence, we term our method ICRA standing for Iterative Concave Rank Approximation. It is further shown that the employed MM approach finds at least a local minimum of the original concave program.

The rest of this paper is organized as follows. After presenting the notations used throughout the paper, in Section \ref{sec:algo}, the main idea and details of the proposed algorithm are described. Section \ref{sec:PAnalysis} gives some theoretical guarantees for the ICRA method as well as a theorem proving the convergence of the exploited optimization scheme. In Section \ref{sec:proofs}, the proofs of theorems and lemmas are presented. In Section \ref{sec:NumExp}, some empirical results from the ICRA method are presented, and it is compared against SRF \cite{MaleBAJ14}, NNM, and reweighted NNM \cite{FazeHB03}. Section \ref{sec:Con} concludes the paper.

\emph{Notations}: For any $\Xb \in \DimDef$, $\nDef$, $\sigma_i(\Xb)$ denotes the $i$th largest singular value, $\Sigb(\Xb) = (\sigma_1(\Xb),\ldots,\sigma_n(\Xb))^T$, and $\|\Xb\|_* \triangleq \sumn \sigma_i(\Xb)$ is the nuclear norm. Besides, it is always assumed that singular values of matrices are sorted in descending order. $\vect(\Xb)$ denotes the vector in $\Rbb^{n_1 n_2}$ with the columns of $\Xb$ stacked on top of one another. $\Sbb^n$ and $\Sbb^n_+$ are used to denote the sets of symmetric and positive semidefinite $n \times n$ real matrices, respectively. For any $\Yb \in \Sbb^{n}$, $\lambda_i(\Yb)$ designates the $i$th largest eigenvalue in magnitude, $\LamVec(\Yb) = \LamVecD(\Yb) = (\lambda_1(\Yb),\ldots,\lambda_n(\Yb))^T$ is the vector of eigenvalues of $\Yb$, and $\trace(\Yb) = \sumn \lambda_i(\Yb)$. Also, $\LamVecA(\Yb) = (\lambda_n(\Yb),\ldots,\lambda_1(\Yb))^T$ denotes the vector of eigenvalues of $\Yb$ in ascending order. For $\Yb,\Zb \in \Sbb^{n}$, $\Yb \succeq \Zb$ and $\Yb \succ \Zb$ means $\Yb - \Zb$ is positive semidefinite and positive definite, respectively. Let $\langle \Xb,\Yb \rangle \triangleq \trace(\Xb^T\Yb)$ and $\langle \xb,\yb \rangle \triangleq \xb^T\yb$ be the inner products on matrix and vector spaces, respectively. As a result, $\|\Xb\|_F \triangleq \langle \Xb,\Xb \rangle^{\frac{1}{2}} = \sqrt{\sum_{i=1}^{n} \sigma_i^2(\Xb)}$ denotes the Frobenius norm, and $\|\xb\|_2 \triangleq \langle \xb,\xb \rangle^{\frac{1}{2}}$ stands for the Euclidean norm. Moreover, $\| \xb \|_{\infty} \triangleq \max_{i} |x_i|$ designates the maximum norm. $\lceil x \rceil$ denotes the smallest integer greater than or equal to $x$. $\Ib_n$ is the identity matrix of order $n$. For a linear operator $\LOp$, let $\mathcal{N}(\ACal) \triangleq \{\Xb \in \DimDef | \ACal(\Xb) = \Zerb\}$.

\section{The ICRA Algorithm} \label{sec:algo}
\subsection{Introduction}
Let
\begin{equation*}
u(x) = \left\{ \,
\begin{IEEEeqnarraybox}[][c]{l?s}
\IEEEstrut
1 & if $x > 0$, \\
0 & if $x = 0$.
\IEEEstrut
\end{IEEEeqnarraybox}
\right.
\end{equation*}
denote the unit step function for $x \geq 0$ so that the rank of a matrix $\Xb$ equals to $\sumn u(\sigma_i(\Xb))$. As $u(x)$ is discontinuous and nondifferentiable, direct minimization of rank is very hard, and all available exact optimizers have doubly exponential complexity \cite{ChistG84}. Consequently, one approach to solve \eqref{RM} is to approximate the unit step function with a suitable one-variable function $f(x)$ and minimize $F(\Xb) = \sumn f(\sigma_i(\Xb))$ as an approximation of the rank function. Herein, for the sake of brevity, we refer to the one- and matrix-variable functions $f(x)$ and $F(\Xb)$ as unit step approximating (UA) and rank approximating (RA) functions, respectively.

\begin{figure}[tb]
\centering
\includegraphics[width=0.49\textwidth]{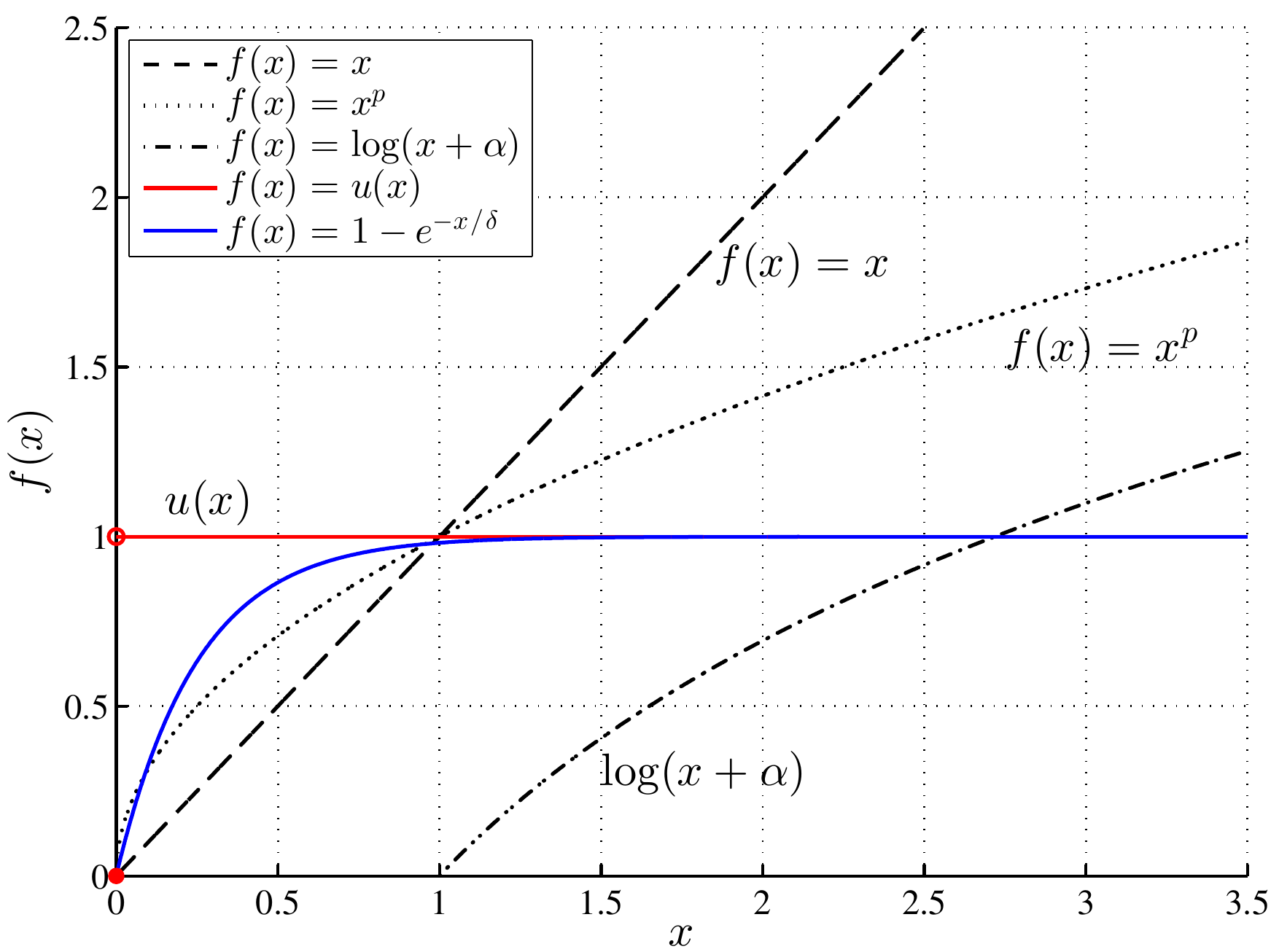}
\vspace{-0.6cm}
\caption{It is known that $\rank(\Xb) = \sumn u(\sigma_i(\Xb))$. Therefore, approximation of the rank function can be converted to the problem of approximating $u(x)$. Different functions used in the literature of rank minimization to approximate the unit step and some of them are plotted in this figure. Among them, $f(x) = 1 - e^{-x/\delta}$ closely matches $u(x)$.} \label{fig:FuncFig}
\end{figure}

Implicitly or explicitly, different one-variable functions have been used to approximate $u(x)$ in some of the existing rank minimization methods. Figure \ref{fig:FuncFig} illustrates some of the available options for approximating the unit step as well as one of the functions used in this work. In this plot, $f(x)=x$ has the worst fitting, though, it leads to nuclear norm minimization, which is the tightest convex relaxation of \eqref{RM} \cite{RechFP10}. $f(x) = x^p, 0 < p < 1,$ which is closer to $u(x)$ yields Schatten-$p$ quasi-norm minimization \cite{MaleBS14}. In \cite{MaleBS14}, theoretically, it is shown that finding the global solution of constrained Schatten-$p$ quasi-norm minimization outperforms NNM. Moreover, experimental observations show superiority of this method to NNM \cite{MohaF12,MarjS12}. $f(x) = \log(x+\alpha)$, in which $\alpha$ is some small constant to ensure positivity of the argument of $\log(\cdot)$, also, results in better performance in recovering low-rank matrices in numerical simulations \cite{FazeHB03}.

Having the above theoretical and experimental results in mind, we expect that finer approximations will give rise to higher performance in recovery of low-rank matrices. Accordingly, we propose using other UA functions like $f(x) = 1-e^{-x/\delta}$ that closely match $u(x)$ for small values of $\delta$. Obviously, $f(x) = 1-e^{-x/\delta}$ is the best approximation among the functions depicted in Figure \ref{fig:FuncFig} in the sense that $\int_{0}^{\infty} |f(t) - u(t)|^2 \mathrm{d}t = \delta / 2$, for every $\delta > 0$, is finite. Furthermore, by this choice, one can control the merit of the approximation by adjusting the parameter $\delta$.

\subsection{The main idea}
Let $\Fdel(\Xb) = h_{\delta}(\SigbXb) = \sumn \fdel(\sigma_i(\Xb))$ denote the rank approximating function. We replace the original ARM problem with
\begin{equation} \label{Fmin}
\min_{\Xb} \Big( \Fdel(\Xb) = \sumn \fdel(\sigma_i(\Xb) \Big) \quad \text{s.t.} \quad \ACal(\Xb)=\bb \text{.}
\end{equation}

When $\delta$ is small, $u(x)$ is well approximated by $\fdel(x)$. However, in this case, $\Fdel(\Xb)$ has many local minima. In contrast, while a larger $\delta$ causes smoother $\Fdel(\Xb)$ with poor approximation quality, $\Fdel(\Xb)$ has smaller number of local minima. In fact, it will be shown in Theorem \ref{InitThm} that when $\delta \rightarrow \infty$, $\delta \Fdel$ converts to a convex function. Consequently, to decrease the chance of getting trapped in local minima while minimizing $\Fdel(\Xb)$, instead of initially minimizing it with a small $\delta$, the ICRA algorithm starts with a large value of $\delta$ ($\delta \rightarrow \infty$). Next, the value of $\delta$ is decreased gradually and the solution of the previous iteration is used as an initial point for minimizing $\Fdel(\Xb)$ at the current iteration with a new $\delta$. Furthermore, we impose the class of functions \{$\fdel$\} to be continuous with respect to $\delta$. From this continuity, we expect that the minimizers of \eqref{Fmin} for successive iterations, let say for $\delta = \delta_i$ and $\delta_{i+1}$, are close to each other as $\delta$ decreases gradually and $\delta_{i+1}$ is in the vicinity of $\delta_{i}$.
Thus, it is more likely that a global minimizer of $\Fdel$ is found. This technique which is known as Graduated NonConvexity (GNC) \cite{BlakZ87} is used in \cite{MaleBAJ14} to solve the affine rank minimization problem.

\subsection{Properties of $\fdel(\cdot)$}
To efficiently solve \eqref{Fmin}, we are interested in differentiable RA functions. The following proposition, which is originally from \cite[Cor.~2.5]{Lewi95}, characterizes the gradient of $\Fdel(\Xb)$ in terms of the derivative of $\fdel(\cdot)$.

\newtheorem{Prop1}{Proposition}
\begin{Prop1} \label{BadGradThm}
Assume that $F : \DimDef \rightarrow \Rbb$ is represented as $F(\Xb)=h\big(\SigbXb\big)$. Let $\Xb = \mathbf{U} \text{diag}(\SigbXb)\mathbf{V}^T$ denote the Singular Value Decomposition (SVD) of $\Xb$. If $h$ is absolutely symmetric\footnote{$h(\xb)$ is absolutely symmetric if it is invariant under arbitrary permutations and sign changes of the components of $\xb$.}, then the subdifferential of $F(\Xb)$ at $\Xb$ is
\begin{equation*}
\partial F(\Xb) = \{ \Ub \diag(\boldsymbol{\theta}) \Vb^T | \boldsymbol{\theta} \in \partial h(\SigbXb) \},
\end{equation*}
where $\partial h(\SigbXb)$ denotes the subdifferential of $h$ at $\SigbXb$.
\end{Prop1}

\smallskip

Clearly, under assumptions of Proposition \ref{BadGradThm}, $\fdel(\cdot)$ must be an even function. This requirement as well as other properties of UA functions cause $\fdel(\cdot)$ to be nondifferentiable at the origin. Therefore, $\Fdel(\Xb)$ becomes nondifferentiable too. This can be seen in another way. Assuming $n_1 \leq n_2$ and $\Xb \Xb^T = \Ub \diag(\lambda_1, \cdots, \lambda_{n_1} ) \Ub^T$ denoting the EigenValue Decomposition (EVD) of $\Xb \Xb^T$, $\fdel(x) = 1 - e^{-x/\delta}$ induces
\begin{equation*}
\Fdel(\Xb) = \trace(I_{n_1} - e^{-(\Xb \Xb^T)^{1/2}/\delta}),
\end{equation*}
 in which $(\Xb \Xb^T)^{1/2} = \Ub \diag(\lambda_1^{1/2}, \cdots, \lambda_{n_1}^{1/2}) \Ub^T$. This reveals that $\Fdel(\Xb)$ is not differentiable at any non full-rank matrix. Nevertheless, if the domain of $\Fdel(\cdot)$ is restricted to the cone of positive semidefinite matrices, we can ignore the requirement that $\fdel(\cdot)$ is symmetric and find concave and differentiable approximations for the rank using the following propositions.\footnote{Propositions \ref{ConcavityThm} and \ref{GradThm} can be restated under the milder condition of $\Yb \in \Sbbn$. However, as our approximation for symmetric matrices relies on the magnitude of eigenvalues, this less restrictive assumption imposes the UA function to be even, making it again nondifferentiable at the origin.}

\newtheorem{Prop2}[Prop1]{Proposition}
\begin{Prop2} \label{ConcavityThm}
Assume that $F : \Sbbp^{n} \rightarrow \Rbb$ is represented as $F(\Yb)=h \big(\LambYb\big)=h \circ \LambYb$. If $h: \Rbb^n \rightarrow \Rbb$ is symmetric and concave, then $F(\Yb)$ is concave.
\begin{proof}
The proof follows from \cite[Cor.~2.7]{Lewi96}.
\end{proof}
\end{Prop2}

\newtheorem{Prop3}[Prop1]{Proposition}
\begin{Prop3} \label{GradThm}
Suppose that $F : \Sbbp^{n} \rightarrow \Rbb$ is represented as $F(\Yb)=h\big(\LambYb\big)= \sumn f(\lambda_i(\Yb))$, where $\Yb \in \Sbbp^{n}$ with the EVD $\Yb = \mathbf{Q} \diag(\LambYb)\mathbf{Q}^T$, $h:\Rbb^{n} \rightarrow \Rbb$ and $f : \Rbb \rightarrow \Rbb$ is differentiable and concave. Then the gradient of $F(\Yb)$ at $\Yb$ is
\begin{equation} \label{GradDef}
\frac{\partial{F(\Yb)}}{\partial{\Yb}}=\mathbf{Q} \text{diag}(\boldsymbol{\theta}) \mathbf{Q}^T,
\end{equation}
where $\boldsymbol{\theta} = \nabla h(\LambYb)$ denotes the gradient of $h$ at $\LambYb$.
\begin{proof}
In~\cite[Thm.~3.2]{Lewi96}, it is shown that if a function $h$ is symmetric and the matrix $\Yb \in \Sbbp^{n}$ has $\LambYb$ in the domain of $h$, then the subdifferential of $F$ is given by
\begin{equation}
\partial \big(h \circ \LambYb \big) = \{ \mathbf{Q} \text{diag}(\boldsymbol{\theta}) \mathbf{Q}^T | \boldsymbol{\theta} \in \partial h \big(\LambYb\big)\}.
\end{equation}
Since $h \big(\LambYb\big)=\sumn f \big(\lambda_i(\Yb)\big)$ is differentiable at $\LambYb$, $\partial h\big(\LambXb\big)$ is a singleton and consequently $\partial \big(h \circ \LambYb \big)$ becomes a singleton. For a convex (concave) function, the subdifferential is singleton if and only if the function is differentiable~\cite{Rock70}. This implies that $F(\Yb)$ is differentiable at $\Yb$ with the above gradient.
\end{proof}
\end{Prop3}

Proposition \ref{GradThm} relaxes the differentiability conditions of Proposition \ref{BadGradThm} by restricting the domain of $\Fdel(\cdot)$. However, we will show in the following subsection that problem \eqref{Fmin} can be converted to another problem in which the argument of $\Fdel(\cdot)$ is positive semidefinite. Putting all the required properties of $\fdel(\cdot)$ together, we are interested in a certain family of UA functions possessing the following property.

\newtheorem{Assump1}{Property}
\begin{Assump1} \label{Assump1}
Let  $f: \Rbb \rightarrow \Rbb$ and define $\fdel(x) \triangleq f(x/\delta)$ for any $\delta > 0$. The class $\{\fdel\}$ is said to possess Property~\ref{Assump1}, if
\begin{enumerate}
\item[(a)] $f$ is real analytic on $(x_0,\infty)$ for some $x_0 < 0$,

\item[(b)] $f$ is strictly concave for $x \geq 0$\footnote{For the most of analysis presented in this paper, concavity of $f(\cdot)$ is sufficient, and strict concavity is merely needed to show that the used optimization algorithm converges to a local minimum.} and concave on $\Rbb$,

\item[(c)] $f(x) = 0 \Leftrightarrow x = 0$,

\item[(d)] for $x \geq 0, f(x)$ is nondecreasing,

\item[(e)] $\lim_{x\rightarrow +\infty}f(x) = 1$.

\end{enumerate}
\end{Assump1}

It follows immediately from Property \ref{Assump1} that, for $x \geq 0$, $\{\fdel\}$ converges pointwise to the unit step function as $\delta\rightarrow 0^{+}$; i.e.,
\begin{equation}\label{eq:fdelLim}
\lim_{\delta \to 0^{+}} \fdel(x)= \left\{
	\begin{array}{rl}
	1 & \text{if } x > 0,\\
	0 & \text{if } x = 0.
	\end{array} \right.
\end{equation}

In addition to UA function $f(x) = 1-e^{-x}$ which is mainly used in this paper, there are other functions that satisfy conditions of Property \ref{Assump1}. For example,
\begin{equation*}
f(x) = \left\{
    \begin{array}{ll}
	\dfrac{x}{x+1} &  x \geq x_0,\\
	- \infty & \text{otherwise},
	\end{array} \right.
\end{equation*}
for some $-1 < x_0 < 0$.

\subsection{Optimization of $\Fdel(\cdot)$ for a specific $\delta$}
The following lemma from \cite{FazeHB03} shows that the original ARM problem is equivalent to
\begin{equation} \label{RMPos}
\min_{(\Xb,\Yb,\Zb)} \rank(\Yb) + \rank(\Zb) ~\text{s.t.}~ \ACal(\Xb) = \bb, \begin{bmatrix} \Yb & \Xb \\ \Xb^T & \Zb \end{bmatrix} \succeq \Zerb,
\end{equation}
where $\Yb \in \Sbb^{n_1}$ and $\Zb \in \Sbb^{n_2}$.

\newtheorem{Lem1}{Lemma}
\begin{Lem1}[{{\hspace{-0.04em}\cite[Lem.~1]{FazeHB03}}}] \label{SDPEmbd}
Let $\Xb \in \DimDef$ be any arbitrary matrix. Then $\rank(\Xb) \leq r$ if and only if there exist matrices $\Yb \in \Sbb^{n_1}$ and $\Zb \in \Sbb^{n_2}$ such that
\begin{equation*}
\rank(\Yb) + \rank(\Zb) \leq 2r, \quad \begin{bmatrix} \Yb & \Xb \\ \Xb^T & \Zb \end{bmatrix} \succeq \Zerb.
\end{equation*}
\end{Lem1}

$\left( \begin{smallmatrix} \Yb & \Xb \\ \Xb^T & \Zb \end{smallmatrix} \right) \succeq \Zerb$ implies that $\Yb \succeq \Zerb, \Zb \succeq \Zerb$ \cite{HornJ91}. Therefore, if $\rank(\Yb) + \rank(\Zb)$ is approximated by
\begin{equation*}
\Fdel(\Yb) + \Fdel(\Zb) = \sum_{i=1}^{n_1} \fdel(\lambda_i(\Yb)) + \sum_{i=1}^{n_2} \fdel(\lambda_i(\Zb)),
\end{equation*}
then, according to Propositions \ref{ConcavityThm} and \ref{GradThm}, $\Fdel(\Yb)$ and $\Fdel(\Zb)$ have the desirable concavity and differentiability properties.

As a result, to extend \eqref{Fmin} to arbitrary matrices with a differentiable and concave RA function,
\begin{equation} \label{FminPos}
\begin{aligned}
& \hspace{-0.1cm} \min_{(\Xb,\Yb,\Zb)}
& & \Fdel(\Yb) + \Fdel(\Zb) \\
& \hspace{-0.01cm} \text{subject to}
& &  \ACal(\Xb) = \bb,~\left( \begin{smallmatrix} \Yb & \Xb \\ \Xb^T & \Zb \end{smallmatrix} \right) \succeq \Zerb, \\
\end{aligned}
\end{equation}
is solved to find a solution to \eqref{Fmin}. A similar approach has been exploited in \cite{FazeHB03} to convert \eqref{Fmin} for $\fdel(x) = \log(x + \alpha)$ to\footnote{For this case, $\fdel(\cdot)$ does not scale with $\delta$.}
\begin{equation} \label{LogPos}
\begin{aligned}
& \hspace{-0.1cm} \min_{(\Xb,\Yb,\Zb)}
& & \log(\det(\Yb + \alpha \Ib_{n_1})) + \log(\det(\Zb + \alpha \Ib_{n_2})) \\
& \hspace{-0.01cm} \text{subject to}
& & \ACal(\Xb) = \bb,~ \left( \begin{smallmatrix} \Yb & \Xb \\ \Xb^T & \Zb \end{smallmatrix} \right) \succeq \Zerb. \\
\end{aligned}
\end{equation}

To solve \eqref{FminPos}, we use a Majorize-Minimize (MM) technique \cite{HuntL04}. In MM approach, the original cost function is replaced with a surrogate function having the following properties. For a vector function $h(\xb) : \Rbb^n \rightarrow \Rbb$, $H(\xb,\tilde{\xb}): \Rbb^n \times \Rbb^n \rightarrow \Rbb$ is a surrogate function at $\tilde{\xb}$ if $H(\xb,\tilde{\xb})$ satisfies
\begin{align*}
H(\tilde{\xb},\tilde{\xb}) & = h(\tilde{\xb}), \\
H(\xb,\tilde{\xb}) & \geq h(\xb), \text{ for all } \xb.
\end{align*}
$H(\xb,\tilde{\xb})$ is also known as tangent-majorant, as the surface $\xb \mapsto H(\xb,\tilde{\xb})$ is tangent to the surface $h(\xb)$ at $\tilde{\xb}$ and lies above it at other points. The underlying idea of MM is to iteratively minimize the surrogate function instead of minimizing the original cost function. More precisely, let $\xb_k$ denote the solution at the $k$th iteration, then $\xb_{k+1}$ is obtained by minimizing the surrogate function at $\xb_k$; that is,
\begin{equation*}
\xb_{k+1} \in \argmin_{\xb \in \FCal} H(\xb,\xb_k),
\end{equation*}
where $\FCal$ denotes the feasible set of the optimization problem. It can be easily shown that $h(\xb_{k+1}) \leq h(\xb_{k})$ proving that the original cost function is continuously decreasing. Naturally, a good choice for a surrogate function is a convex one which can be easily optimized. In our problem, since $\Fdel(\Yb)$ is concave, the first-order concavity condition implies that
\begin{equation*}
\Fdel(\Yb) \leq \Fdel(\tilde{\Yb}) + \langle \Yb - \tilde{\Yb}, \nabla \Fdel(\tilde{\Yb}) \rangle,
\end{equation*}
for some $\tilde{\Yb}$ in the feasible set. As a result, $\Hdel(\Yb,\Yb_k) \triangleq \Fdel(\Yb_k) + \langle \Yb - \Yb_k, \nabla \Fdel(\Yb_k) \rangle$ is chosen as a surrogate function for $\Fdel(\Yb)$. With a tiny abuse of notation, let, likewise, $\Hdel(\Zb,\Zb_k) \triangleq \Fdel(\Zb_k) + \langle \Zb - \Zb_k, \nabla \Fdel(\Zb_k) \rangle$ denote the surrogate function for $\Fdel(\Zb)$. Applying the MM approach, problem \eqref{FminPos}, for a fixed $\delta$, can be optimized by iteratively solving
\begin{IEEEeqnarray}{lll} \label{FminPosMM}
&& (\Xb_{k+1},\Yb_{k+1},\Zb_{k+1}) = \nonumber \\
&& \begin{aligned}
& \hspace{-0.1cm} \argmin_{(\Xb,\Yb,\Zb)}
& & \langle \nabla \Fdel(\Yb_k), \Yb \rangle  + \langle \nabla \Fdel(\Zb_k), \Zb \rangle \\
& \hspace{-0.01cm} \text{subject to}
& & \ACal(\Xb) = \bb,~\left( \begin{smallmatrix} \Yb & \Xb \\ \Xb^T & \Zb \end{smallmatrix} \right) \succeq \Zerb, \\
\end{aligned}
\end{IEEEeqnarray}
until convergence. It is easy to verify that the above program is SDP, and it is shown in Section \ref{sec:LConv} that it converges to a local minimum of \eqref{FminPos}.

\subsection{Initialization}
As pointed out earlier, in the GNC procedure, we initially solve \eqref{Fmin} or \eqref{FminPos} for $\delta$ tending to $\infty$. In this case, as shown in the following theorem, whose proof is given in Section \ref{sec:InitThmSec}, \eqref{Fmin} and \eqref{FminPos} can be optimized by solving \eqref{NNM}.

\newtheorem{Thm1}{Theorem}
\begin{Thm1} \label{InitThm}
For any class of functions $\{\fdel\}$ possessing Property~\ref{Assump1} and any $\Xb \in \DimDef, \Yb \in \Sbb^{n_1},\Zb \in \Sbb^{n_2}$,
\begin{IEEEeqnarray*}{rCl} 
\lim_{\delta \to \infty} \frac{\delta}{\gamma} \Fdel(\Xb) & =  & \| \Xb \|_*, \\
\lim_{\delta \to \infty} \frac{\delta}{\gamma} \big( \Fdel(\Yb) + \Fdel(\Zb) \big) & =  & \trace(\Yb) + \trace(\Zb),
\end{IEEEeqnarray*}
where $\gamma = f'(0) \neq 0$. Furthermore, \begin{IEEEeqnarray*}{rCl}
\IEEEeqnarraymulticol{3}{l}{\lim_{\delta \to \infty} \argmin_{\Xb} \{\Fdel(\Xb) | \ACal(\Xb) = \bb \} } \\ \qquad\qquad\qquad\qquad\qquad
& = & \argmin_{\Xb} \{\|\Xb\|_* | \ACal(\Xb) = \bb \},
\end{IEEEeqnarray*}
provided that NNM has a unique solution.
\end{Thm1}

\medskip

A solution to \eqref{NNM} can be obtained by optimizing the following equivalent problem \cite{RechFP10}
\begin{equation} \label{NNMPos}
\begin{aligned}
& \hspace{-0.1cm} \min_{(\Xb,\Yb,\Zb)}
& & \trace(\Yb) + \trace(\Zb) \\
& \hspace{-0.01cm} \text{subject to}
& &  \ACal(\Xb) = \bb,~ \left( \begin{smallmatrix} \Yb & \Xb \\ \Xb^T & \Zb \end{smallmatrix} \right) \succeq \Zerb. \\
\end{aligned}
\end{equation}
Accordingly, $\Xb_0,\Yb_0,\Zb_0$ are initialized by solving \eqref{NNMPos}.

\subsection{The final algorithm}
Applying all the introduced stages of the algorithm to the UA function $\fdel(x) = 1 - e^{-x/\delta}$, the ICRA algorithm is summarized in Figure \ref{fig:Algo}. In addition, the following remarks give complementary comments about implementation details by describing algorithm parameters and their selection rules.

\begin{figure}
\centering
\begin{lrbox}{\ieeealgbox}
\begin{minipage}{\dimexpr\columnwidth-2\fboxsep-2\fboxrule}
\begin{algorithmic}
\STATE \hspace{-0.4cm} Input: $\ACal(\cdot), \bb, \fdel(\cdot)$
\STATE \hspace{-0.4cm} Initialization:
\end{algorithmic}
\begin{algorithmic}[1]
\STATE $\Xb_0 = \argmin_{\Xb} \{ \|\Xb\|_* | \ACal(\Xb) = \bb \}$.
\STATE $\delta_0 = 8 \sigma_1(\Xb_0)$.
\STATE $c$: decreasing factor for $\delta$.
\STATE $\epsilon_1$, $\epsilon_2$: stopping thresholds for main and internal loops.
\end{algorithmic}
\begin{algorithmic}
\STATE \hspace{-0.4cm} Body:
\end{algorithmic}
\begin{algorithmic}[1]
\STATE $i = 0, \delta = \delta_i$.
\WHILE{$d_1 > \epsilon_1$}
\STATE $j = 0, \Xbh_0 = \Xb_i$.
    \STATE \textbf{while} $d_2 > \epsilon_2$ \textbf{do}
    \STATE \hspace{\algorithmicindent}  \vspace{-1.5em}
                                        \begin{IEEEeqnarray*}{lll}
                                        && \hspace{-1.3em} (\Xbh_{j+1},\Ybh_{j+1},\Zbh_{j+1}) = \\
                                        && \hspace{-1.3em} \begin{aligned}
                                        & \hspace{-0.1cm} \argmin_{(\Xb,\Yb,\Zb)}
                                        & & \langle \nabla \Fdel(\Ybh_j), \Yb \rangle + \langle \nabla \Fdel(\Zbh_j), \Zb \rangle \\
                                        & \hspace{-0.01cm} \text{subject to}
                                        & & \ACal(\Xb) = \bb,~\left( \begin{smallmatrix} \Yb & \Xb \\ \Xb^T & \Zb \end{smallmatrix} \right) \succeq \Zerb. \\
                                        \end{aligned}
                                        \end{IEEEeqnarray*}
    \STATE \hspace{\algorithmicindent} $d_2 = \| \Xbh_{j+1}  - \Xbh_{j} \|_F / \| \Xbh_{j} \|_F$.
    \STATE \hspace{\algorithmicindent} $j = j + 1$.
    \STATE \textbf{end while}
\STATE $\Xb_{i+1} = \Xbh_{j}$.
\STATE $d_1 = \| \Xb_{i+1}  - \Xb_{i} \|_F / \| \Xb_{i} \|_F$.
\STATE $i = i + 1, \delta=c \delta$.

\ENDWHILE
\end{algorithmic}
\begin{algorithmic}
\STATE \hspace{-0.4cm} Output: $\Xb_i$
\end{algorithmic}
\end{minipage}
\end{lrbox}\noindent\fbox{\usebox{\ieeealgbox}}
\caption{The ICRA Algorithm.}  \label{fig:Algo}
\end{figure}

\emph{Remark 1.} As depicted in Figure \ref{fig:Algo}, $\delta$ is updated as $\delta_i = c \delta_{i-1}$ for $i \geq 1$. We will examine how to choose a suitable decreasing factor $c$ in Section \ref{sec:NumExp} in more details, yet $c \in (0.1,0.5)$ is a good choice in general. Furthermore, $\delta_0$ is set to $8 \sigma_1(\Xb_0)$ because it is easy to verify that $1-e^{-\sigma_i(\Xb_0) / \delta_0}$ is closely approximated by $\sigma_i(\Xb_0) / \delta_0$ with this choice of $\delta_0$. Hence, this $\delta_0$ acts as if it tends to $\infty$.

\emph{Remark 2.} $d_1 = \| \Xb_{i+1}  - \Xb_{i} \|_F / \| \Xb_{i} \|_F$ and $d_2 = \| \Xbh_{j+1}  - \Xbh_{j} \|_F / \| \Xbh_{j} \|_F$, as measures of distances between results of successive iterations, are used to stop execution of the external and internal loops, respectively. Moreover, $\epsilon_1$ and $\epsilon_2$ are usually set to $10^{-2}$ to settle down $\Xb_{i+1}$ and $\Xbh_{j+1}$ to vicinity of 1\% distance of the previous solutions $\Xb_{i}$  and $\Xbh_{j}$.

\emph{Remark 3.} For $\fdel(x) = 1 - e^{-x/\delta}$, the gradient of $\Fdel(\Ybh_j)$ and $\Fdel(\Zbh_j)$ are given by
\begin{align*}
\Fdel(\Ybh_j) & = \frac{1}{\delta} ~ \Pb \diag(e^{-\lambda_1(\Ybh_j)/\delta },\cdots,e^{-\lambda_{n_1}(\Ybh_j)/\delta}) \Pb^T,\\
\Fdel(\Zbh_j) & = \frac{1}{\delta} ~ \Qb \diag(e^{-\lambda_1(\Zbh_j)/\delta },\cdots,e^{-\lambda_{n_2}(\Zbh_j)/\delta}) \Qb^T,
\end{align*}
where $\Pb \diag(\LamVec(\Ybh_j)) \Pb^T$ and $\Qb \diag(\LamVec(\Zbh_j)) \Qb^T$ denote the EVD of $\Ybh_j$ and $\Zbh_j$, respectively.

\emph{Remark 4.} Following the same argument as in \cite{MohaF10b}, problem \eqref{FminPosMM} can be cast as a re weighted nuclear norm minimization; i.e.,
\begin{equation*}
\Xb_{k+1} = \argmin \| \Wb_k^{l} \Xb \Wb_k^{r} \|_* ~~\text{s.t.}~~ \ACal(\Xb) = \bb.
\end{equation*}
If $\Ub \SigMat \Vb^T$ denotes the SVD of $\Wb_k^{l} \Xb_{k+1} \Wb_k^{r}$, then weighting matrices as well as $\Yb_{k+1},\Zb_{k+1}$ are updated by
\begin{IEEEeqnarray*}{rCl}
\Yb_{k+1} & = & \big( \Wb_k^l \big)^{-1} \Ub \SigMat \Ub^T \big( \Wb_k^l \big)^{-1},\\
\Zb_{k+1} & = & \big( \Wb_k^r \big)^{-1} \Vb \SigMat \Vb^T \big( \Wb_k^r \big)^{-1},\\
\Wb_{k+1}^l & = & \Big( \nabla \Fdel(\Yb_{k+1}) \Big)^{\frac{1}{2}}, \Wb_{k+1}^r = \Big( \nabla \Fdel(\Zb_{k+1})\Big)^{\frac{1}{2}}.
\end{IEEEeqnarray*}
There are efficient solvers for the NNM like FPCA \cite{MaGC11} and APG \cite{TohY10}. As a result, one can exploit these algorithm to solve \eqref{FminPosMM} more efficiently than SDP.

\emph{Remark 5.} \eqref{Fmin} can be generalized to the following setting for taking into account the noise in measurements
\begin{equation} \label{FminNoisy}
\min_{\Xb} \Fdel(\Xb) ~ \text{subject to} ~ \| \ACal(\Xb)-\bb \|_2 \leq \epsilon.
\end{equation}
Consequently, the following program can be solved instead of \eqref{FminPosMM}
\begin{IEEEeqnarray}{lll}
&& (\Xb_{k+1},\Yb_{k+1},\Zb_{k+1}) = \nonumber \\
&& \begin{aligned}
& \hspace{-0.1cm} \argmin_{(\Xb,\Yb,\Zb)}
& & \langle \nabla \Fdel(\Yb_k), \Yb \rangle  + \langle \nabla \Fdel(\Zb_k), \Zb \rangle \\
& \hspace{-0.01cm} \text{subject to}
& & \| \ACal(\Xb) - \bb \|_2 \leq \epsilon,~\left( \begin{smallmatrix} \Yb & \Xb \\ \Xb^T & \Zb \end{smallmatrix} \right) \succeq \Zerb.
\end{aligned}
\end{IEEEeqnarray}

\section{Performance Analysis} \label{sec:PAnalysis}
In this section, we analyze the performance of the ICRA algorithm in recovery of low-rank matrices. First, in Section \ref{sec:Uniq}, a necessary and sufficient condition for exact recovery of \eqref{Fmin} is presented. The sufficient condition is based on null-space properties of the measurement operator. Next, exploiting results established in \cite{MaleBAJ14}, in Section \ref{sec:RConv}, we prove that the sequence of minimizers of \eqref{Fmin}, for a decreasing sequence of $\delta$, converges to the minimum rank solution. We will not discuss the issue of global convergence; instead, it is shown that if the MM approach is applied, program \eqref{FminPosMM} converges, at least, to a local minimizer of \eqref{FminPos}.

\subsection{Uniqueness} \label{sec:Uniq}
One simple way to characterize the conditions under which a method can successfully find the exact solution in both sparse vector and low-rank matrix recovery from underdetermined linear measurements is to use null-space properties of the measurement operator. In the vector case, for a general function inducing a `sparsity measure', a necessary and sufficient condition for exact recovery is derived in \cite{GribN07}. Here, we generalize some results of \cite{GribN07} to low-rank matrix recovery and introduce a necessary and sufficient condition for the success of \eqref{Fmin}. Furthermore, it is shown that global optimization of \eqref{Fmin} uniquely recovers matrices of higher or equal ranks than those of uniquely recoverable by NNM. The proof of the following lemmas and theorem are given in Section \ref{sec:UniqProofs}.

The results of the next two lemmas are valid for not only $\fdel(x) = f(x/\delta)$ in \eqref{Fmin} but also any $f:\Rbb \to \Rbb$ which is used in
\begin{equation*}
\min_{\Xb} \Big( F(\Xb) = \sumn f(\sigma_i(\Xb) \Big) \quad \text{subject to} \quad \ACal(\Xb)=\bb
\end{equation*}
to recover a low-rank matrix.

\newtheorem{Lem2}[Lem1]{Lemma}
\begin{Lem2} \label{NSPSuf}
Every matrix $\Xb \in \DimDef$ of rank at most $r$ can be uniquely recovered using \eqref{Fmin} for any $f$ possessing Property \ref{Assump1}-(b) to \ref{Assump1}-(d), if, $\forall \Wb \in \NullDefmZ$,
\begin{equation*}
\sum_{i=1}^{r} f\big(\sigma_i(\Wb)\big) < \sum_{i=r+1}^{n} f\big(\sigma_i(\Wb)\big).
\end{equation*}
\end{Lem2}

In general, extending Lemma \ref{NSPSuf} to the noisy rank minimization is not straight-forward. In fact, even in the vector case, robust recovery conditions (RRC)\footnote{The so-called RRC guarantees stable recovery of sparse vectors from noisy measurements using minimization of a sparsity measure inducing function.} for a sparsity measure have been derived only for the $\ell_p$ quasi-norm \cite{LiuJG13_2}. Nevertheless, a recent work \cite{LiuJG13_2} proves that, under some mild assumptions, the sets of measurement matrices satisfying exact recovery conditions and RRC differ by a set of measure zero. Accordingly, recalling the strong parallels between RM and $\ell_0$-minimization \cite{RechFP10}, roughly speaking, we expect that under the same conditions as in Lemma \ref{NSPSuf}, \eqref{FminNoisy} can recover matrices close to the solutions of \eqref{RMNoisy} in the Frobenius-norm sense.

\newtheorem{Lem3}[Lem1]{Lemma}
\begin{Lem3} \label{NSPNec}
Under the same assumptions on $f$ as in Lemma \ref{NSPSuf}, if, for some $\Wb \in \NullDefmZ$,
\begin{equation} \label{SufIneq}
\sum_{i=1}^{r} f\big(\sigma_i(\Wb)\big) \geq \sum_{i=r+1}^{n} f\big(\sigma_i(\Wb)\big),
\end{equation}
then there exist $\Xb$ and $\Xb'$ such that $\rank(\Xb) \leq r,\ACal(\Xb)=\ACal(\Xb')$ and $F(\Xb') \leq F(\Xb)$.
\end{Lem3}

The sufficient condition in Lemma \ref{NSPSuf} can be also described by the following inequality
\begin{equation*}
2 \sum_{i=1}^{r} f\big(\sigma_i(\Wb)\big) < \sum_{i=1}^{n} f\big(\sigma_i(\Wb)\big).
\end{equation*}
As a result, if we define
\begin{equation*}
\theta_{f}(r,\ACal) \triangleq \sup_{\Wb \in \NullDefmZ} \frac{\sum_{i=1}^{r} f(\sigma_i(\Wb))}{\sum_{i=1}^{n} f(\sigma_i(\Wb))},
\end{equation*}
the uniqueness can be characterized as: \emph{All matrices of rank at most $r$ are uniquely recovered by \eqref{Fmin} if $\theta_{f}(r,\ACal) < 1/2$}. In fact, $\theta_{f}$ extends a similar parameter defined in \cite{GribN07} for $\ell_0$-norm minimization.

Let $r_{\fdel}^*(\ACal)$ denote the maximum rank such that all matrices $\Xb$ with $\rank(\Xb) \leq r_{\fdel}^*(\ACal)$ can be uniquely recovered by \eqref{Fmin}. In particular, $r_{\text{arm}}^*(\ACal)$ and $r_{\text{nnm}}^*(\ACal)$ are the corresponding values for $\fdel(x)=u(x)$ and $\fdel(x)=x$; that is, original rank minimization problem, \eqref{RM}, and nuclear norm minimization, \eqref{NNM}. Then we have the following result.

\newtheorem{Thm6}[Thm1]{Theorem}
\begin{Thm6} \label{Sup}
For any $\fdel(\cdot)$ possessing Property \ref{Assump1},
\begin{equation*}
r_{\text{nnm}}^*(\ACal) \leq r_{\fdel}^*(\ACal) \leq r_{\text{arm}}^*(\ACal).
\end{equation*}
\end{Thm6}

\subsection{Convergence to the rank function} \label{sec:RConv}
The following definition, which like $\theta_{f}(r,\ACal)$ depends on the null space of $\ACal$, is used to show that when $\delta \to 0$, the solution of \eqref{Fmin} tends toward the minimum rank solution of \eqref{RM}. In other words, in order to get arbitrarily close to the minimum rank solution, it is sufficient to solve \eqref{Fmin} for a properly chosen $\delta$ which depends on the employed UA function.

\newtheorem{Def1}{Definition}
\begin{Def1}[Spherical Section Property~\cite{MohaFH11,DvijF10}]
The linear operator $\ACal$ possesses $\Delta$-spherical section property if, for all $\Wb \in \NCal(\ACal) \setminus \{\Zerb\}$, $\|\Wb\|_*^2/\|\Wb\|_F^2 \geq \Delta(\ACal)$. In other words, spherical section constant of the linear operator $\ACal$ is defined as
\begin{equation*}
\Delta(\ACal) \triangleq \min_{\Wb \in \NCal(\ACal) \setminus \{\Zerb\}} \frac{\|\Wb\|_*^2}{\|\Wb\|_F^2}.
\end{equation*}
\end{Def1}

The following proposition is originally from \cite[Thm.~4]{MaleBAJ14}. Although different assumptions were imposed on the UA functions in the proof of \cite{MaleBAJ14}, the authors merely used properties that are common to our assumptions, making the result applicable also to our analysis.

\newtheorem{Prop4}[Prop1]{Proposition}
\begin{Prop4}\label{ConvRankThm}
Assume $\ACal$ has $\Delta$-spherical property and $\{\fdel\}$ possesses Property~\ref{Assump1}. Let $\Xb_0$ be the unique solution to \eqref{RM} and let $\Xbd$ denote a solution to \eqref{Fmin}. Then
\begin{equation*}
\|\Xbd - \Xb_0\|_F \leq \frac{n \alpha_{\delta}}{\sqrt{\Delta}-\sqrt{\UDelmOne}},
\end{equation*}
where $\alpha_{\delta} = \big| \fdel^{-1}(1-\frac{1}{n}) \big|$, and, consequently,
\begin{equation*}
\lim_{\delta \to 0^{+}} \Xb_{\delta} = \Xb_{0}.
\end{equation*}
\end{Prop4}

This result is of particular interest since the best result available for NNM shows that if $\rank(\Xb) < \Delta / 4$, then $\Xb$ can be uniquely recovered \cite{MohaFH11} which is more restrictive than $\rank(\Xb) < \Delta / 2$, a sufficient condition for the uniqueness of the solution of \eqref{RM}. However, the above proposition proves that we can find accurate estimate of the original solution whether it is recoverable by NNM or not.

\subsection{Convergence Analysis} \label{sec:LConv}
The next theorem whose proof is left to Section \ref{sec:LConvSec} proves that the MM approach proposed in \eqref{FminPosMM} to solve \eqref{FminPos} will find a local minimizer of \eqref{FminPos}.

\newtheorem{Thm9}[Thm1]{Theorem}
\begin{Thm9} \label{LConvThm}
The sequence of $\{(\Xb_k,\Yb_k,\Zb_k)\}$ is convergent to a local minimizer of \eqref{FminPos}.
\end{Thm9}

\section{Proofs} \label{sec:proofs}
\subsection{Proof of Theorem \ref{InitThm}} \label{sec:InitThmSec}
\begin{proof}
Using the Taylor expansion, $f(\cdot)$ can be formulated as
\begin{equation*}
f(s)=\gamma s+g(s),
\end{equation*}
where $\gamma=f'(0)$ and
\begin{equation} \label{glim}
\lim_{s \to 0} \frac{g(s)}{s}=0.
\end{equation}
$\gamma$ cannot be 0 because the first-order concavity condition implies that, for any $ x > 0$,
\begin{equation*}
f(x) \leq f(0) + x f'(0) = \gamma x,
\end{equation*}
and $\gamma = 0$ converts the above inequality to $f(x) \leq 0$ which contradicts Property \ref{Assump1}. Now, $\Fdel(\cdot)$ can be represented as
\begin{IEEEeqnarray}{rCl} \label{FTaylor}
\Fdel(\Xb) & = & \sumn \fdel\big(\sigma_i(\Xb)\big) \nonumber \\
& = &\frac{\gamma}{\delta}\|\Xb\|_* + \sumn g(\sigma_i(\Xb)/\delta).
\end{IEEEeqnarray}
\eqref{FTaylor} can be reformulated as
\begin{equation} \label{Taylor}
\frac{\delta}{\gamma} \Fdel(\Xb) = \|\Xb\|_* + \frac{1}{\gamma} \sumn \sigma_i(\Xb) \frac{g(\sigma_i(\Xb)/\delta)}{\sigma_i(\Xb)/\delta}.
\end{equation}
By virtue of \eqref{glim}, it follows that
\begin{equation*}
\lim_{\delta \to \infty} \frac{\delta}{\gamma} \Fdel(\Xb) = \| \Xb \|_*.
\end{equation*}
Following the same line of argument, it can be easily verified
\begin{equation*}
\lim_{\delta \to \infty} \frac{\delta}{\gamma} \big( \Fdel(\Yb) + \Fdel(\Zb) \big) =  \trace(\Yb) + \trace(\Zb).
\end{equation*}
To prove the second part, let
\begin{IEEEeqnarray*}{llCl}
& \Xbh & = & \argmin_{\Xb} \{\|\Xb\|_* | \ACal(\Xb) = \bb \}, \\
& \Xbd & = & \argmin_{\Xb} \{\Fdel(\Xb) | \ACal(\Xb) = \bb \}.
\end{IEEEeqnarray*}
From \eqref{Taylor} and the inequality
\begin{equation*}
\Big | \sumn x_i y_i \Big | \leq \Big( \sumn \big|x_i\big| \Big) \Big( \sumn \big|y_i\big| \Big),
\end{equation*}
we have
\begin{IEEEeqnarray*}{rCl}
\delta \Fdel(\Xb) & \leq  & \| \Xb \|_* \Big (\gamma + \sumn \frac{|g(\sigma_i(\Xb)/\delta)|}{\sigma_i(\Xb)/\delta} \Big),\\
\delta \Fdel(\Xb) & \geq  & \| \Xb \|_* \Big (\gamma - \sumn \frac{|g(\sigma_i(\Xb)/\delta)|}{\sigma_i(\Xb)/\delta} \Big).
\end{IEEEeqnarray*}
The above inequalities as well as \eqref{glim} imply that $\forall \epsilon > 0, \exists \delta_0,$ such that $\forall \delta > \delta_0$
\begin{equation*}
\gamma - \epsilon \leq \frac{\delta \Fdel(\Xb)}{\| \Xb \|_*} \leq \gamma + \epsilon.
\end{equation*}
$\Xbd$ is a solution to \eqref{Fmin}, so $\delta \Fdel(\Xbd) \leq \delta \Fdel(\Xbh)$. Furthermore, we have $\| \Xbh \|_* \leq \|\Xbd\|_*$ since $\Xbh$ is the unique solutio of \eqref{NNM}. Therefore, for $\epsilon < \gamma$, we obtain
\begin{equation*}
(\gamma - \epsilon) \| \Xbh \|_* \! \leq \!\! (\gamma - \epsilon) \| \Xbd \|_* \! \leq \!  \delta \Fdel(\Xbd) \!\! \leq  \! \delta \Fdel(\Xbh) \!\! \leq \!\! (\gamma + \epsilon) \| \Xbh \|_*
\end{equation*}
which proves that $\lim_{\delta \to \infty} \|\Xbd\|_* = \|\Xbh\|_*$. As $\Xbh$ is the unique solution to \eqref{NNM} (under the same equality constraints), it can be concluded that $\lim_{\delta \to \infty} \Xbd = \Xbh$.
\end{proof}

\subsection{Proofs of Propositions \ref{NSPSuf} and \ref{NSPNec} and Theorem \ref{Sup}} \label{sec:UniqProofs}
Before proofs, we need the following definition, corollary, and lemmas.

\newtheorem{Def2}[Def1]{Definition}
\begin{Def2}[\hspace{-0.05em}\cite{HornJ90}]
A function $\Phi(\xb) : \Rbb^n \rightarrow \Rbb$ is called symmetric gauge if it is a norm on $\Rbb^n$ and absolutely symmetric.
\end{Def2}

\newtheorem{Lem6}[Lem1]{Lemma}
\begin{Lem6}[{{\hspace{-0.04em}\cite[Cor.~2.3]{ZhanQ10}}}] \label{SubAddLem}
Let $\Phi$ be a symmetric gauge function and $f: [0,\infty) \rightarrow [0,\infty)$ be a concave function with $f(0)=0$. Then for $\Ab,\Bb \in \DimDef$,
\begin{equation*}
\Phi \Big( f\big(\Sigb(\Ab)\big) - f\big(\Sigb(\Bb)\big) \Big) \leq \Phi \Big( f \big( \Sigb(\Ab - \Bb) \big) \Big),
\end{equation*}
where $f\big(\xb) = (f(x_1), \ldots, f(x_n))^T$.
\end{Lem6}

\newtheorem{Lemma11}[Lem1]{Lemma}
\begin{Lemma11} \label{GrowthLemma}
For any function possessing Property~\ref{Assump1}, $f(x)/x$ is nonincreasing for $x > 0$.
\begin{proof}
Let $g(x) = f(x)/x$. It is sufficient to show that $g'(x)=\big(xf'(x)-f(x)\big)/ x^2$ is nonpositive for $x > 0$. $f(x)$ is concave, so we can write
\begin{equation*}
f(0) \leq f(x) + (0-x)f'(x)
\end{equation*}
for any $x > 0$ which proves that $g'(x) \leq 0$.
\end{proof}
\end{Lemma11}

\newtheorem{Cor1}{Corollary}
\begin{Cor1} \label{SubAddCor}
Let $\Ab,\Bb \in \DimDef$. For any $f$ possessing Property \ref{Assump1}-(b) to \ref{Assump1}-(d),
\begin{equation} \label{mainInEq}
\sumn f\big(\sigma_i(\Ab - \Bb)\big) \geq \sumn \big|f\big(\sigma_i(\Ab)\big) - f\big(\sigma_i(\Bb)\big)\big|.
\end{equation}
\begin{proof}
$\Phi(\xb) = \sum_{i=1}^n |x_i|$ and $f(\cdot)$ satisfy conditions of Lemma \ref{SubAddLem}; thus, \eqref{mainInEq} immediately follows.
\end{proof}
\end{Cor1}

\begin{proof}[Proof of Lemma \ref{NSPSuf}]
The proof is similar to \cite[Lem.~6]{OymaH10} and extends uniqueness condition from NNM to a larger class of functions possessing Property \ref{Assump1}-(b) to \ref{Assump1}-(d). Assuming $\ACal(\Xb) = \bb$, all feasible solutions to \eqref{Fmin} can be formulated as $\Xb + \Wb$ for some $\Wb \in \NullDef$. To show that $\Xb$ is a unique solution to \eqref{Fmin}, it is sufficient to prove that, $\forall \Wb \in \NullDefmZ$, $F(\Xb + \Wb) > F(\Xb)$. Starting from Corollary \ref{SubAddCor}, we can write that
\begin{IEEEeqnarray*}{rCl}
F(\Xb + \Wb) \! & = & \! \sum_{i=1}^{n} f \big( \sigma_i(\Xb + \Wb) \big) \\
& \geq & \! \sum_{i=1}^{n} \big|f(\sigma_i(\Xb)) \!-\! f(\sigma_i(\Wb))\big| \\
& = & \! \sum_{i=1}^{r} \big|f(\sigma_i(\Xb)) \!-\! f(\sigma_i(\Wb))\big| \! + \!\!\!\! \sum_{i=r+1}^{n} \!\!\! f(\sigma_i(\Wb)) \\
& \geq & \! \sum_{i=1}^{r} f(\sigma_i(\Xb)) \!-\! f(\sigma_i(\Wb)) \! + \!\!\!\! \sum_{i=r+1}^{n} \!\!\! f(\sigma_i(\Wb)) \\
& > & \! \sum_{i=1}^{r} f(\sigma_i(\Xb)) = F(\Xb),
\end{IEEEeqnarray*}
which completes the proof.
\end{proof}

\begin{proof}[Proof of Lemma \ref{NSPNec}]
Let
\begin{equation*}
\Wb = \Ub \diag(\sigma_1,\ldots,\sigma_n) \Vb^T
\end{equation*}
denote the SVD of $\Wb$. Choose
\begin{IEEEeqnarray*}{rCrll}
\Xb & = & -\Ub & \diag(\sigma_1,\ldots,\sigma_r, 0, \ldots,0) & \Vb^T \\
\Xb' & = & \Ub & \diag(0,\ldots,0, \sigma_{r+1}, \ldots,\sigma_{n}) & \Vb^T.
\end{IEEEeqnarray*}
Obviously, $\Wb = \Xb' - \Xb$, $\ACal(\Xb)=\ACal(\Xb')$, and $\rank(\Xb) \leq r$. On the other hand, \eqref{SufIneq} implies that
\begin{equation*}
F(\Xb')=\sum_{i=r+1}^{n} f(\sigma_i(\Wb)) \leq \sum_{i=1}^{r} f(\sigma_i(\Wb)) = F(\Xb).
\end{equation*}
\end{proof}

\begin{proof}[Proof of Theorem \ref{Sup}]
Lemma \ref{GrowthLemma} implies that, for $x > 0$, $\fdel(x)/x$  is nonincreasing. Hence, following a similar argument as in \cite[Thm.~5]{GribN07}, one can easily verify that, for any $\Wb \neq \Zerb$, $\sum_{i=1}^{r} \sigma_i(\Wb) / \sum_{i=1}^{r} \fdel( \sigma_i(\Wb) )$ is a nonincreasing sequence in $r$. Consequently,
\begin{equation*}
\frac{\sum_{i=1}^{n} \sigma_i(\Wb)}{\sum_{i=1}^{n} \fdel(\sigma_i(\Wb))} \leq \frac{\sum_{i=1}^{r} \sigma_i(\Wb)}{\sum_{i=1}^{r} \fdel(\sigma_i(\Wb))},
\end{equation*}
or,
\begin{equation*}
\frac{\sum_{i=1}^{r} \fdel(\sigma_i(\Wb))}{\sum_{i=1}^{n} \fdel(\sigma_i(\Wb))} \leq \frac{\sum_{i=1}^{r} \sigma_i(\Wb)}{\sum_{i=1}^{n} \sigma_i(\Wb)},
\end{equation*}
which shows $\theta_{\fdel}(r,\ACal) \leq \theta_{\text{nnm}}(r,\ACal)$ for any $r \leq n$. $\theta_{\fdel}(r,\ACal),\theta_{\text{nnm}}(r,\ACal)$ are increasing in $r$, so it can be concluded that $r_{\fdel}^*(\ACal) \geq r_{\text{nnm}}^*(\ACal)$. Similarly, it can be shown that $\sum_{i=1}^{r} u(\sigma_i(\Wb)) / \sum_{i=1}^{r} \fdel( \sigma_i(\Wb) )$ is a nondecreasing sequence, and
\begin{equation*}
\frac{\sum_{i=1}^{r} u(\sigma_i(\Wb))}{\sum_{i=1}^{n} u(\sigma_i(\Wb))} \leq \frac{\sum_{i=1}^{r} \fdel(\sigma_i(\Wb))}{\sum_{i=1}^{n} \fdel(\sigma_i(\Wb))},
\end{equation*}
confirming that $r_{\text{arm}}^*(\ACal) \geq r_{\fdel}^*(\ACal)$.
\end{proof}

\subsection{Proof of Theorem \ref{LConvThm}} \label{sec:LConvSec}
We start with the following lemmas. The first lemma is originally from \cite[Lem.~II.1]{Lass95}.

\newtheorem{Lem4}[Lem1]{Lemma}
\begin{Lem4}[{{\hspace{-0.04em}\cite[Lem.~II.1]{Lass95}}}] \label{TrIneq}
Let $\Ab,\Bb \in \Sbb^n$; then
\begin{equation*}
\sumn \lambda_{n-i+1}(\Ab) \lambda_{i}(\Bb) \leq \trace(\Ab\Bb) \leq \sumn \lambda_{i}(\Ab) \lambda_{i}(\Bb).
\end{equation*}
\end{Lem4}

\newtheorem{Lem5}[Lem1]{Lemma}
\begin{Lem5} \label{CurvatureLem1}
Assume that $F : \Sbbn \rightarrow \Rbb$ is represented as $F(\Xb)=h\big(\LambXb\big)=\sum_{i=1}^{n} f(\lambda(\Xb))$ in which $f: \Rbb \rightarrow \Rbb$. If $f(\cdot)$ is twice differentiable and strictly concave, then $F(\Xb)$ is strictly concave, and there is some $m > 0$ such that, for any bounded $\Xb,\Yb \in \Sbbn$, $\Xb \neq \Yb$,
\begin{equation} \label{IneqMat1}
F(\Yb) - F(\Xb) \leq \langle \Yb - \Xb , \nabla F(\Xb) \rangle -\frac{m}{2} \| \Yb - \Xb \|_F^2.
\end{equation}
\begin{proof}
First, it is shown that $F(\cdot)$ is strictly concave, then \eqref{IneqMat1} follows as a result. To this end, notice that strict concavity of $f(\cdot)$ implies that $h(\cdot)$ is strictly concave too. From the first-order concavity condition, it is known that $h$ is strictly concave if and only if, for any $\xb \neq \yb$,
\begin{equation*}
h(\yb) < h(\xb) + \langle \yb - \xb, \nabla h(\xb) \rangle.
\end{equation*}
Propositions \ref{ConcavityThm} and \ref{GradThm} together imply that $F(\cdot)$ is differentiable. Thus, substituting $\xb,\yb$ with $\LamVecD(\Xb),\LamVecD(\Yb)$ in the above inequality gives
\begin{equation} \label{FIneq}
F(\Yb) < F(\Xb) + \langle \LamVecD(\Yb) - \LamVecD(\Xb), \nabla h(\LamVecD(\Xb)) \rangle.
\end{equation}
Let $\Xb = \Ub \diag(\LamVecD(\Xb)) \Ub^T$ denote the EVD of $\Xb$. Applying Proposition \ref{GradThm} on $F(\cdot)$ yields
\begin{align*}
\nabla F(\Xb) & = \Ub \diag(f'(\lambda_1(\Xb)),\cdots,f'(\lambda_n(\Xb))) \Ub^T \\
              & = \Ub \diag(\nabla h(\LamVecD(\Xb))) \Ub^T.
\end{align*}
Therefore,
\begin{align} \label{InnerProEq}
\langle \Xb, \nabla F(\Xb) \rangle & = \trace( \diag(\LamVecD(\Xb)) \diag(\nabla h(\LamVecD(\Xb))) ) \nonumber \\
& = \langle \LamVecD(\Xb) ,\nabla h(\LamVecD(\Xb)) \rangle.
\end{align}
Also,
\begin{equation*}
\langle \Yb, \nabla F(\Xb) \rangle = \trace(\Yb \nabla F(\Xb)) \overset{a}{\geq} \langle \LamVecD(\Yb), \LamVecA(\nabla F(\Xb)) \rangle,
\end{equation*}
where (a) follows from Lemma \ref{TrIneq}. Since $f(\cdot)$ is strictly concave, $f'(\cdot)$ is decreasing and $f'(\lambda_i(\Xb)) \geq f'(\lambda_j(\Xb))$ for $i \geq j$. Therefore, $\LamVecA(\nabla F(\Xb)) = \nabla h(\LamVecD(\Xb))$, and the above inequality becomes
\begin{equation} \label{InnerPrIneq1}
\langle \Yb, \nabla F(\Xb) \rangle \geq \langle \LamVecD(\Yb), \nabla h(\LamVecD(\Xb)).
\end{equation}
Substituting \eqref{InnerProEq} and \eqref{InnerPrIneq1} in \eqref{FIneq}, we obtain
\begin{equation}
F(\Yb) < F(\Xb) + \langle \Yb - \Xb, \nabla F(\Xb) \rangle,
\end{equation}
which shows that $F(\cdot)$ is strictly concave.

The Hessian of $h(\xb)$ is given by
\begin{equation*}
\nabla^2 h(\xb) = \diag(f''(x_1),\cdots,f''(x_n)).
\end{equation*}
As $f(\cdot)$ is strictly concave, for any bounded $U > 0$, there is a $m' > 0$ such that $f''(x) \leq -m'$ for any $|x| \leq U$, and it follows that $\nabla^2 h(\xb) \preceq -m' \Ib$ for all $\xb$ with $\|\xb\|_{\infty} \leq U$. Further, assuming $\|\xb\|_{\infty},\|\yb\|_{\infty} \leq U$, we have
\begin{equation*}
h(\yb) = h(\xb) + \langle \yb - \xb, \nabla h(\xb) \rangle + \frac{1}{2} (\yb - \xb)^T \nabla^2 h(\zb) (\yb - \xb)
\end{equation*}
for some $\zb$ in the line segment connecting $\xb$ and $\yb$. Using $\nabla^2 h(\zb) \preceq -m' \Ib$, we get
\begin{equation*}
h(\yb) \leq h(\xb) + \langle \yb - \xb, \nabla h(\xb) \rangle - \frac{m'}{2} \| \yb - \xb \|_2^2.
\end{equation*}
Similarly, for the function $F(\cdot)$ which is strictly concave, there is some $m > 0$ such that for any bounded $\Xb,\Yb \in \Sbbn, \Xb \neq \Yb$, \begin{equation*}
F(\Yb) - F(\Xb) \leq \langle \Yb - \Xb, \nabla F(\Xb) \rangle - \frac{m}{2} \| \Yb - \Xb \|_F^2,
\end{equation*}
which completes the proof.
\end{proof}
\end{Lem5}

\begin{proof}[Proof of Theorem \ref{LConvThm}]
First, we show that the sequence $\{(\Xb_k,\Yb_k,\Zb_k)\}$ is bounded and convergent. Since $\Fdel(\Yb)$ and $\Fdel(\Zb)$ are concave, we can write that, for every $\Yb \in \Sbbp^{n_1}, \; \Zb \in \Sbbp^{n_2}$,
\begin{IEEEeqnarray*}{rCrCCCl}
\Fdel(\Yb) & \leq & \Fdel(\Yb_k) & + & \langle \Yb - \Yb_k, \nabla \Fdel(\Yb_k) \rangle & = & \Hdel(\Yb,\Yb_k), \\
\Fdel(\Zb) & \leq & \Fdel(\Zb_k) & + & \langle \Zb - \Zb_k, \nabla \Fdel(\Zb_k) \rangle & = & \Hdel(\Zb,\Zb_k).
\end{IEEEeqnarray*}
In the MM step, the next point is updated by
\begin{equation*}
(\Xb_{k+1},\Yb_{k+1},\Zb_{k+1}) =
\begin{aligned}
& \hspace{-0.1cm} \argmin_{(\Xb,\Yb,\Zb)}
& & \!\! \Hdel(\Yb,\Yb_k) + \Hdel(\Zb,\Zb_k) \\
& \hspace{-0.01cm} \text{subject to}
& & \!\! (\Xb,\Yb,\Zb) \in \FCal, \\
\end{aligned}
\end{equation*}
where
\begin{equation*}
\FCal = \{(\Xb,\Yb,\Zb) | \ACal(\Xb) = \bb, \begin{bmatrix} \Yb & \Xb \\ \Xb^T & \Zb \end{bmatrix} \succeq \Zerb \}
\end{equation*}
denotes the feasible set. Clearly,
\begin{equation*}
\Hdel(\Yb_{k+1},\Yb_k) + \Hdel(\Zb_{k+1},\Zb_k) \leq \Hdel(\Yb,\Yb_k) + \Hdel(\Zb,\Zb_k).
\end{equation*}
Therefore, for all $k$,
\begin{align} \label{SeqDec}
\Fdel(\Yb_{k+1}) + \Fdel(\Zb_{k+1}) & \leq \Hdel(\Yb_{k+1},\Yb_k) + \Hdel(\Zb_{k+1},\Zb_k) \nonumber \\
& \leq  \Hdel(\Yb_k,\Yb_k) + \Hdel(\Zb_k,\Zb_k) \nonumber \\
& = \Fdel(\Yb_{k}) + \Fdel(\Zb_{k}).
\end{align}
From \eqref{SeqDec} and $\Fdel(\Yb_k),\Fdel(\Zb_k) \geq 0$, we can conclude that the sequence $\{\Fdel(\Yb_{k}) + \Fdel(\Zb_{k})\}$ is convergent. Assume that \eqref{FminPosMM} is initialized with $(\Xb_0, \Yb_0)$. We have
\begin{equation*}
\Fdel(\Yb_k) + \Fdel(\Zb_k) \leq \Fdel(\Yb_0) + \Fdel(\Zb_0), \quad \forall k \geq 1,
\end{equation*}
showing $\{\Yb_k\}$ and $\{\Zb_k\}$ are bounded. Moreover, from the constraints
\begin{equation*}
\begin{bmatrix} \Yb_k & \Xb_k \\ \Xb_k^T & \Zb_k \end{bmatrix} \succeq \Zerb,
\end{equation*}
\cite[Lem.~3.5.12]{HornJ91} implies that there is a matrix $\Cb$ with $\sigma_1(\Cb) \leq 1$ such that $\Xb_k = \Yb_k^{\frac{1}{2}} \Cb \Zb_k^{\frac{1}{2}}$ proving that $\{\Xb_k\}$ is also bounded. To show that these sequences are convergent too, we start by applying Lemma \ref{CurvatureLem1} on $\Fdel(\Yb)$ and $\Fdel(\Zb)$ to get
\begin{IEEEeqnarray}{rCl} \label{Ineq81}
\frac{m}{2} \| \Yb_{k+1} - \Yb_{k} \|_F^2 & \leq & \Fdel(\Yb_{k}) - \Fdel(\Yb_{k+1}) \nonumber \\
& \quad + & \langle \Yb_{k+1} - \Yb_{k}, \nabla \Fdel(\Yb_{k}) \rangle, \\
\frac{m}{2} \| \Zb_{k+1} - \Zb_{k} \|_F^2 & \leq & \Fdel(\Zb_{k}) - \Fdel(\Zb_{k+1}) \nonumber \\
& \quad + & \langle \Zb_{k+1} - \Zb_{k}, \nabla \Fdel(\Zb_{k}) \rangle.  \label{Ineq82}
\end{IEEEeqnarray}
From \eqref{FminPosMM}, we have
\begin{IEEEeqnarray*}{lll}
&& (\Xb_{k+1},\Yb_{k+1},\Zb_{k+1}) = \nonumber \\
&& \begin{aligned}
& \hspace{-0.1cm} \argmin_{(\Xb,\Yb,\Zb)}
& & \langle \nabla \Fdel(\Yb_k), \Yb \rangle  + \langle \nabla \Fdel(\Zb_k), \Zb \rangle \\
& \hspace{-0.01cm} \text{subject to}
& & (\Xb,\Yb,\Zb) \in \FCal, \\
\end{aligned}
\end{IEEEeqnarray*}
As a consequence,
\begin{IEEEeqnarray*}{rCl}
\langle \nabla \Fdel(\Yb_k),\Yb_{k+1} \rangle + \langle \nabla \Fdel(\Zb_k),\Zb_{k+1} \rangle & \leq & \langle \nabla \Fdel(\Yb_k),\Yb_{k} \rangle \\
& \quad  + & \langle \nabla \Fdel(\Zb_k),\Zb_{k} \rangle,
\end{IEEEeqnarray*}
or,
\begin{equation*}
\langle \Yb_{k+1} - \Yb_{k},\nabla \Fdel(\Yb_k) \rangle + \langle \Zb_{k+1} - \Zb_{k},\nabla \Fdel(\Zb_k) \rangle \leq 0.
\end{equation*}
Combining \eqref{Ineq81} and \eqref{Ineq82} and knowing that $\langle \Yb_{k+1} - \Yb_{k},\nabla \Fdel(\Yb_k) \rangle + \langle \Zb_{k+1} - \Zb_{k},\nabla \Fdel(\Zb_k) \rangle$ is nonpositive, it can be obtained that
\begin{IEEEeqnarray*}{rCl}
\IEEEeqnarraymulticol{3}{l}{\frac{m}{2} \| \Yb_{k+1} - \Yb_{k} \|_F^2  + \frac{m}{2} \| \Zb_{k+1} - \Zb_{k} \|_F^2} \\ \qquad\qquad
& \leq & \Fdel(\Yb_{k}) - \Fdel(\Yb_{k+1}) + \Fdel(\Zb_{k}) - \Fdel(\Zb_{k+1}),
\end{IEEEeqnarray*}
and, consequently,
\begin{IEEEeqnarray}{rCl} \label{IneqFro1}
\frac{m}{2} \| \Yb_{k+1} - \Yb_{k} \|_F^2 & \leq & \Fdel(\Yb_{k}) - \Fdel(\Yb_{k+1}) \nonumber \\
& \quad + & \Fdel(\Zb_{k}) - \Fdel(\Zb_{k+1}), \\
\frac{m}{2} \| \Zb_{k+1} - \Zb_{k} \|_F^2 & \leq & \Fdel(\Yb_{k}) - \Fdel(\Yb_{k+1}) \nonumber \\
& \quad + & \Fdel(\Zb_{k}) - \Fdel(\Zb_{k+1}). \label{IneqFro2}
\end{IEEEeqnarray}
Summing over $k$, it follows from \eqref{IneqFro1} and \eqref{IneqFro2} that
\begin{IEEEeqnarray*}{rCl}
\frac{m}{2} \sum_{k=0}^{N} \| \Yb_{k+1} - \Yb_{k} \|_F^2 & \leq & \Fdel(\Yb_{0}) + \Fdel(\Zb_{0}),\\
\frac{m}{2} \sum_{k=0}^{N} \| \Zb_{k+1} - \Zb_{k} \|_F^2 & \leq & \Fdel(\Yb_{0}) + \Fdel(\Zb_{0}).
\end{IEEEeqnarray*}
This shows that $\frac{m}{2} \sum_{k=0}^{N} \| \Yb_{k+1} - \Yb_{k} \|_F^2$ and $\frac{m}{2} \sum_{k=0}^{N} \| \Zb_{k+1} - \Zb_{k} \|_F^2$ converge when $N \to \infty$, which, in turn, proves that $\{\Yb_k\}$ and $\{\Zb_k\}$ are convergent. Following the same line of argument made to show that $\{\Xb_k\}$ is bounded, convergence of $\{\Xb_k\}$ follows from $\left( \begin{smallmatrix} \Yb_k & \Xb_k \\ \Xb_k^T & \Zb_k \end{smallmatrix} \right) \succeq \Zerb$ and convergence of $\{\Yb_k\},\{\Zb_k\}$. To show that $\{(\Xb_k,\Yb_k,\Zb_k)\}$ converges to a local minimum of \eqref{FminPos}, we cast \eqref{FminPosMM} as a standard SDP. First, note that
\begin{equation*}
\ACal(\Xb) = \bb \Leftrightarrow \langle \Ab_i,\Xb \rangle = b_i , \quad i = 1, \cdots, m,
\end{equation*}
for some $\Ab_i \in \Rbb^{n_1 \times n_2}$. By introducing
\begin{IEEEeqnarray*}{rCl}
\Ab_i' & \triangleq & \begin{bmatrix} \Zerb & \frac{1}{2} \Ab_i \\ \frac{1}{2} \Ab_i^T & \Zerb \end{bmatrix}, \Cb \triangleq \begin{bmatrix} \nabla \Fdel(\Yb_k) & \Zerb \\ \Zerb & \nabla \Fdel(\Zb_k) \end{bmatrix}, \\
\Wb & \triangleq & \begin{bmatrix} \Yb & \Xb \\ \Xb^T & \Zb \end{bmatrix},
\end{IEEEeqnarray*}
\eqref{FminPosMM} converts to
\begin{equation} \label{SDP}
\begin{aligned}
& \hspace{-0.05cm} \min_{\Wb}
& & \trace(\Cb \Wb) \\
& \hspace{-0.01cm} \text{subject to}
& & \trace(\Ab_i' \Wb) = b_i,~i = 1, \cdots, m, \\
&&& \hspace{+0.07cm} \Wb \succeq \Zerb.
\end{aligned}
\end{equation}
Let $\{\Xb_k,\Yb_k,\Zb_k\} \rightarrow \{\Xb^*,\Yb^*,\Zb^*\}$ as $k \rightarrow \infty$. The Karush-Kuhn-Tucker (KKT) conditions for \eqref{SDP} \cite{AntoL07} implies that, $\exists \yb^* \in \Rbb^m, \Sb^* \in \Sbb^{n_1+n_2}$ such that\begin{itemize}
      \item $\sum_{i=1}^{m} y_i^* \Ab_i' + \Sb^* = \begin{bmatrix} \nabla \Fdel(\Yb^*) & \Zerb \\ \Zerb & \nabla \Fdel(\Zb^*) \end{bmatrix}$,
      \item $\trace(\Ab_i' \Wb) = b_i,~ i = 1, \cdots, m$,
      \item $\Sb^* \Wb^* = \Zerb$,
      \item $\Sb^* \succeq \Zerb, \Wb^* \succeq \Zerb$,
    \end{itemize}
where $\Wb^* = \begin{bmatrix} \Yb^* & \Xb^* \\ \Xb^{*T} & \Zb^* \end{bmatrix}$. It can be easily verified that the above conditions are the KKT conditions for the original problem
\begin{equation*}
\begin{aligned}
& \hspace{-0.05cm} \min_{\Wb}
& & \Fdel(\Yb) + \Fdel(\Zb) \\
& \hspace{-0.01cm}\text{subject to}
& & \hspace{-0.07cm} \trace(\Ab_i' \Wb) = b_i,~ i = 1, \cdots, m, \\
&&& \Wb \succeq \Zerb,
\end{aligned}
\end{equation*}
which together with \eqref{SeqDec} and concavity of the cost function confirms that $(\Xb^*,\Yb^*,\Zb^*)$ is a local minimizer of \eqref{FminPos}.
\end{proof}

\section{Numerical Experiments} \label{sec:NumExp}
In this section, we present a numerical evaluation of the performance of the ICRA algorithm. First, the effect of parameter $c$ in the accuracy of recovering low-rank matrices is analyzed. Next, after proposing a suitable choice for $c$, the evolution of the phase transitions of the ICRA algorithm in solving MC and ARM problems when $\delta$ decreases from $\infty$ is illustrated. Finally, superiority of the proposed algorithm in MC and ARM problems is demonstrated via simulations. Toward this end, ICRA is compared to NNM, the method of \cite{FazeHB03}, and SRF which already outperforms some of the state-of-the-art algorithms in the MC problem \cite{MaleBAJ14}. As mentioned earlier, in \cite{FazeHB03}, Fazel et al proposed to replace \eqref{RM} with \eqref{LogPos}. To solve \eqref{LogPos}, they proposed to use a Majorize-Minimize technique which leads to solving the following SDP iteratively,
\begin{IEEEeqnarray*}{lll}
&& (\Xb_{k+1},\Yb_{k+1},\Zb_{k+1}) = \nonumber \\
&& \begin{aligned}
& \hspace{-0.1cm} \argmin_{(\Xb,\Yb,\Zb)}
& & \!\!\!\! \trace((\Yb_k + \alpha \Ib_{n_1})^{-1} \Yb) + \trace((\Zb_k + \alpha \Ib_{n_2})^{-1} \Zb) \\
& \hspace{-0.01cm} \text{s.t.}
& & \hspace{-0.07cm} \!\!\!\! \left( \begin{smallmatrix} \Yb & \Xb \\ \Xb^T & \Zb \end{smallmatrix} \right) \succeq \Zerb,~\ACal(\Xb) = \bb. \\
\end{aligned}
\end{IEEEeqnarray*}
Although appears to constitute an instance of \eqref{FminPos}, for this replacement of the ARM, $f(x)=\log(x+\alpha)$ does not satisfy some of the requirements in Property \ref{Assump1}. This algorithm is referred as LGD (LoG-Determinant) in the sequel.

We use random matrices as solutions to \eqref{RM} and \eqref{MC} and random linear operators in our simulations. In particular, to generate a random matrix $\Xb \in \DimDef$ of rank $r$, $\Xb^{l} \in \Rbb^{n_1 \times r}$ and $\Xb^{r} \in \Rbb^{r \times n_1}$, whose entries are identically and independently distributed (iid) from a zero-mean, unit-variance Gaussian distribution $\ensuremath{N(0,1)}$, are generated. Then we set $\Xb = \Xb^{l} \Xb^{r}$. The constraints $\ACal(\Xb) = \bb$ are converted to $\Ab \vect(\Xb) = \bb$, where $\Ab \in \Rbb^{m \times n_1 n_2}$ is the matrix representation of $\ACal$, and every elements of $\Ab$ is iid from $\ensuremath{N(0,1)}$. Furthermore, in MC scenarios, revealed entries are selected uniformly at random from all the elements of $\Xb$. Let $\Xbh$ designate the output of one of the above algorithms to recover $\Xb$. For measuring the accuracy of reconstruction, we define $\SNR \triangleq 20 \log_{10} (\| \Xb \|_F / \| \Xb - \Xbh \|_F)$ in dB as the reconstruction SNR. Furthermore, $d_r = r(n_1 + n_2 - r)$ denotes the number of degrees of freedom for a real-valued matrix of dimensions $n_1 \times n_2$ with rank $r$ \cite{CandR09}.

In all simulations, square matrices are considered, and $n_1 = n_2 = n$ is always set to 30. Moreover, always, $\epsilon_1$ and $\epsilon_2$ are fixed to $10^{-2}$ to stop both internal and external loops when the current solution changes only 1\% from the previous one. $f(x) = 1 - e^{-x}$ is the UA function in all the following experiments. All simulations are performed in MATLAB  7.14 environment, and CVX \cite{cvx} is used to solve \eqref{FminPosMM}.

\emph{Experiment 1.} The parameter $c$ is used to control the decay rate of $\delta$ in refining the rank approximation. More specifically, at the $i$th iteration of the external loop, $\delta_i$ is set to $c \delta_{i-1}$. The optimal choice of $c$ is a function of the aspects of the problem under consideration. However, roughly speaking, as the number of measurements decreases toward the degrees of freedom of the solution, larger values of $c$ should be chosen. In contrast, in problems with larger ratio of the number of measurements to the degrees of freedom, smaller values of $c$ lead to less number of iterations while $\SNR$ does not degrade considerably.

\begin{figure}[tb]
\centering
\includegraphics[width=0.49\textwidth]{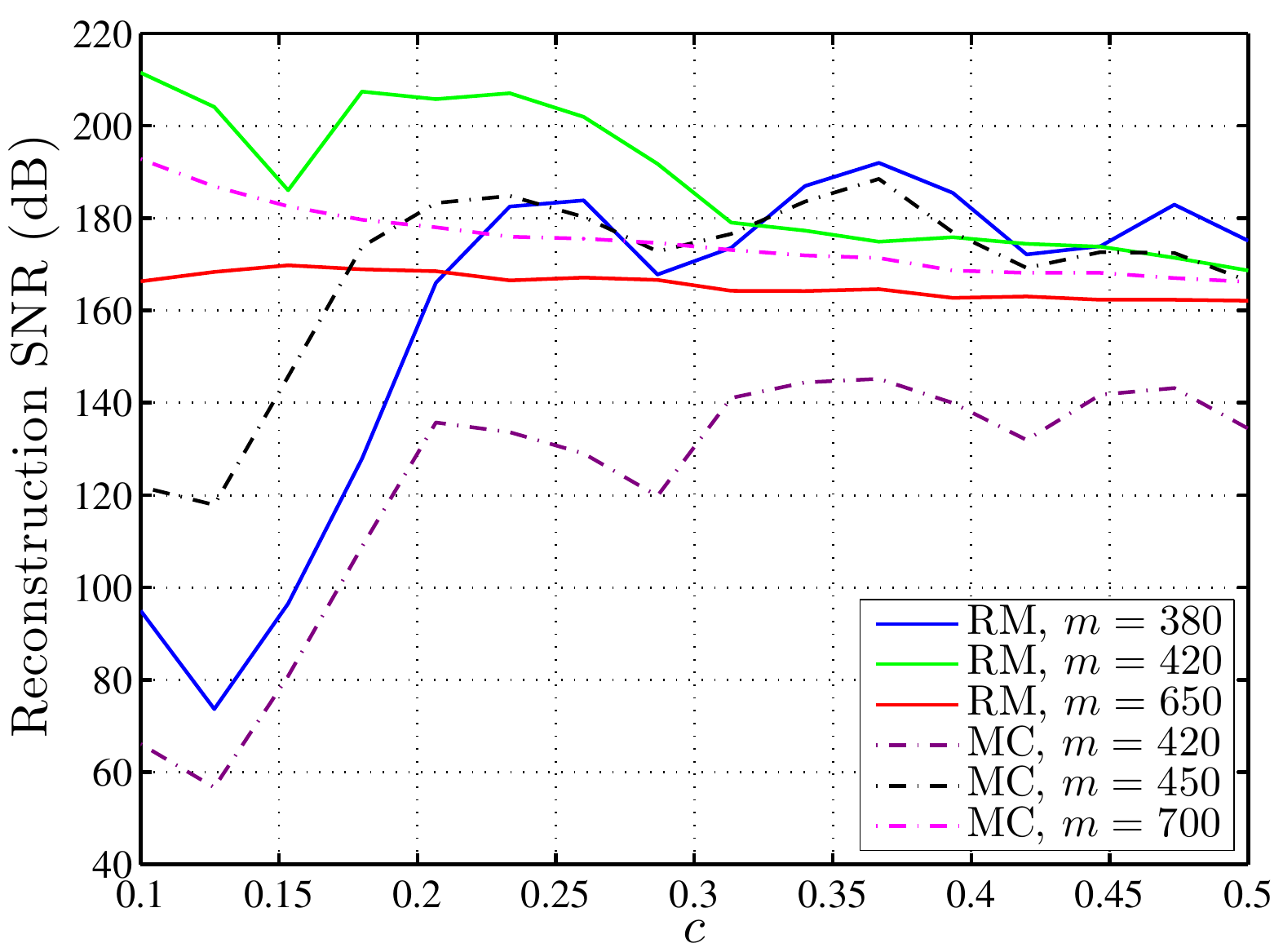}
\vspace{-0.6cm}
\caption{Averaged $\SNR$'s of the ICRA algorithm in solving ARM and MC problems are plotted versus $c$ for 6 different number of measurements. Matrix dimensions are set to $30 \times 30$, and $r$ is fixed to 6 in all simulations. To obtain accurate estimates of the $\SNR$, in each problem, 100 Monte-Carlo simulations are run, and results are averaged.} \label{fig:CFig}
\end{figure}

In this experiment, the above rule is numerically verified. For $30 \times 30$ randomly generated matrices of rank 6, six different ARM and MC problems are solved to cover cases where $m / d_r$ is small or large. $c$ is changed from 0.1 to 0.5. Trials are repeated, for each value of $m$, 100 times, and $\SNR$'s are averaged over these trials. Figure \ref{fig:CFig} shows $\SNR$ as a function of $c$. Clearly, when there is sufficiently large number of measurements, $\SNR$ is approximately independent of $c$. Thus, since increasing $c$ gives rise to more number of iterations, it should be chosen as small as possible. On the other hand, for smaller number of measurements, reconstruction SNR depends on $c$. However, after passing a critical value, $\SNR$ remains approximately unchanged. Therefor, to have the lowest computational complexity, $c$ should be selected a bit above that critical value. Applying the above rule, in the rest of experiments, $c$ is chosen to be 0.2.

\emph{Experiment 2.} This experiment is devoted to analyze the performance of the proposed algorithms as it proceeds with finer approximations of the rank function. To that end, the \emph{phase transition} graph, which similar to the CS framework indicates the region of perfect recovery and failure in solving rank minimization problems \cite{RechFP10,OymaH10}, is utilized. To empirically generate the phase transition graphs, $r$ is changed from 1 to $n$, and, for a fixed $r$, $m$ is swept from $d_r$ to $n^2$. For every pair $(r,m)$, 50 random realizations of $\Xb$ are generated and empirical recovery rates according to the solutions obtained in the initialization step and the next three consecutive iterations of the external loop are calculated. This procedure is run for both ARM and MC settings, and a solution is declared to be recovered if reconstruction SNR is greater than 60 dB.

\begin{figure}[tb]
        \subfigure[]{%
                \label{fig:DevMC1}
                \includegraphics[width=0.23\textwidth]{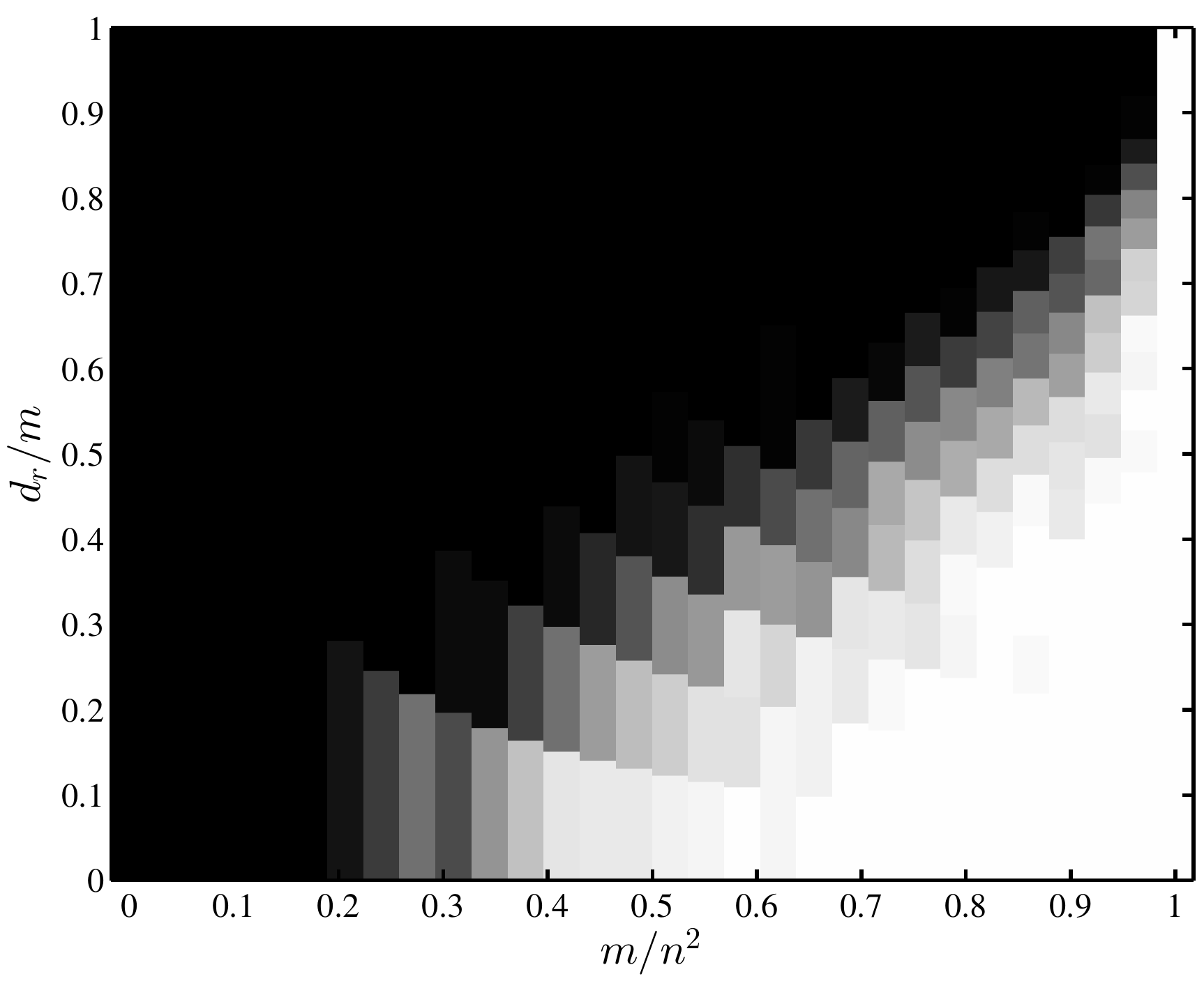}
        }%
        \subfigure[]{%
                \label{fig:DevMC2}
                \includegraphics[width=0.23\textwidth]{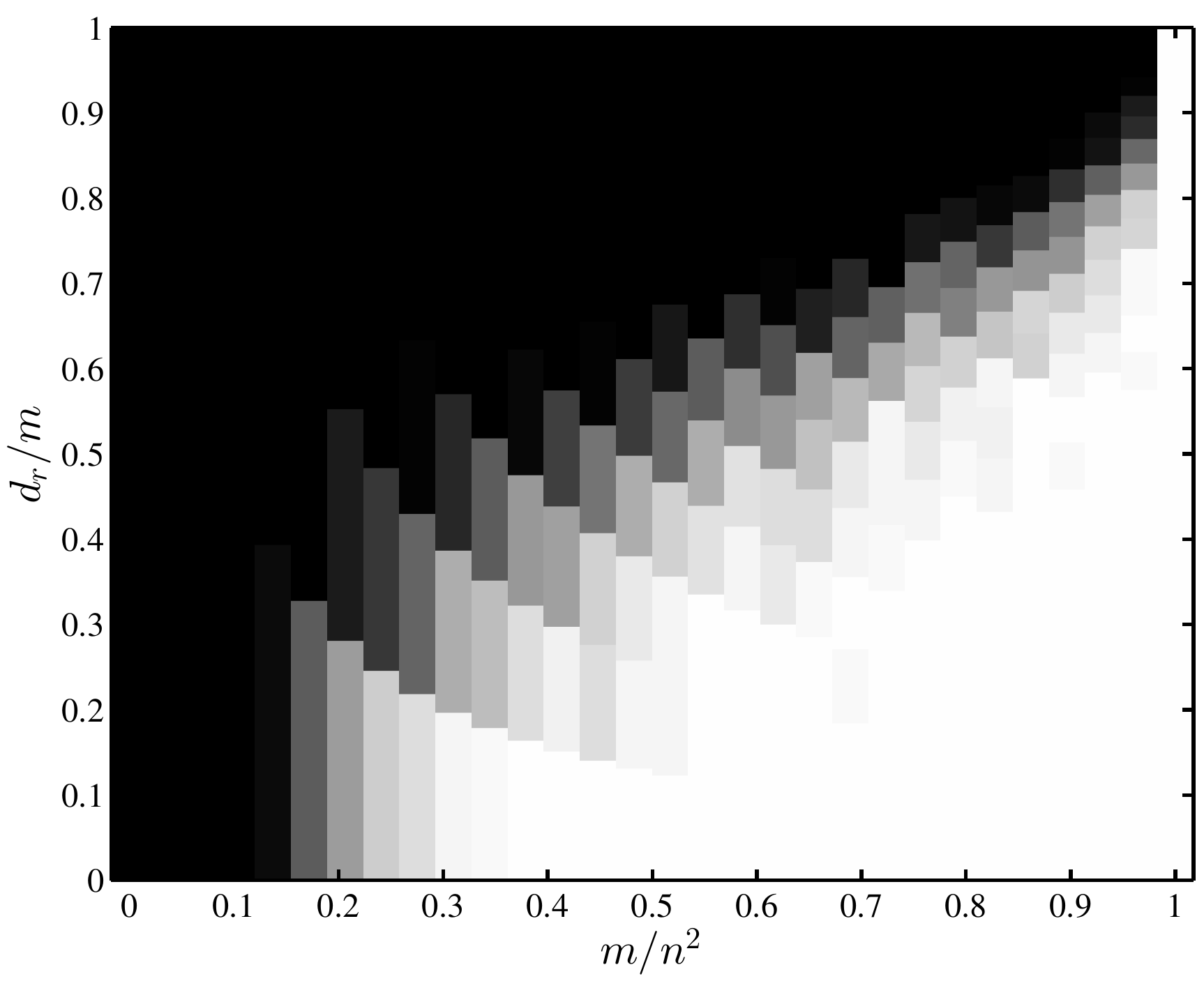}
        }%
        \vspace{-0.35cm}
        \subfigure[]{%
                \label{fig:DevMC3}
                \includegraphics[width=0.23\textwidth]{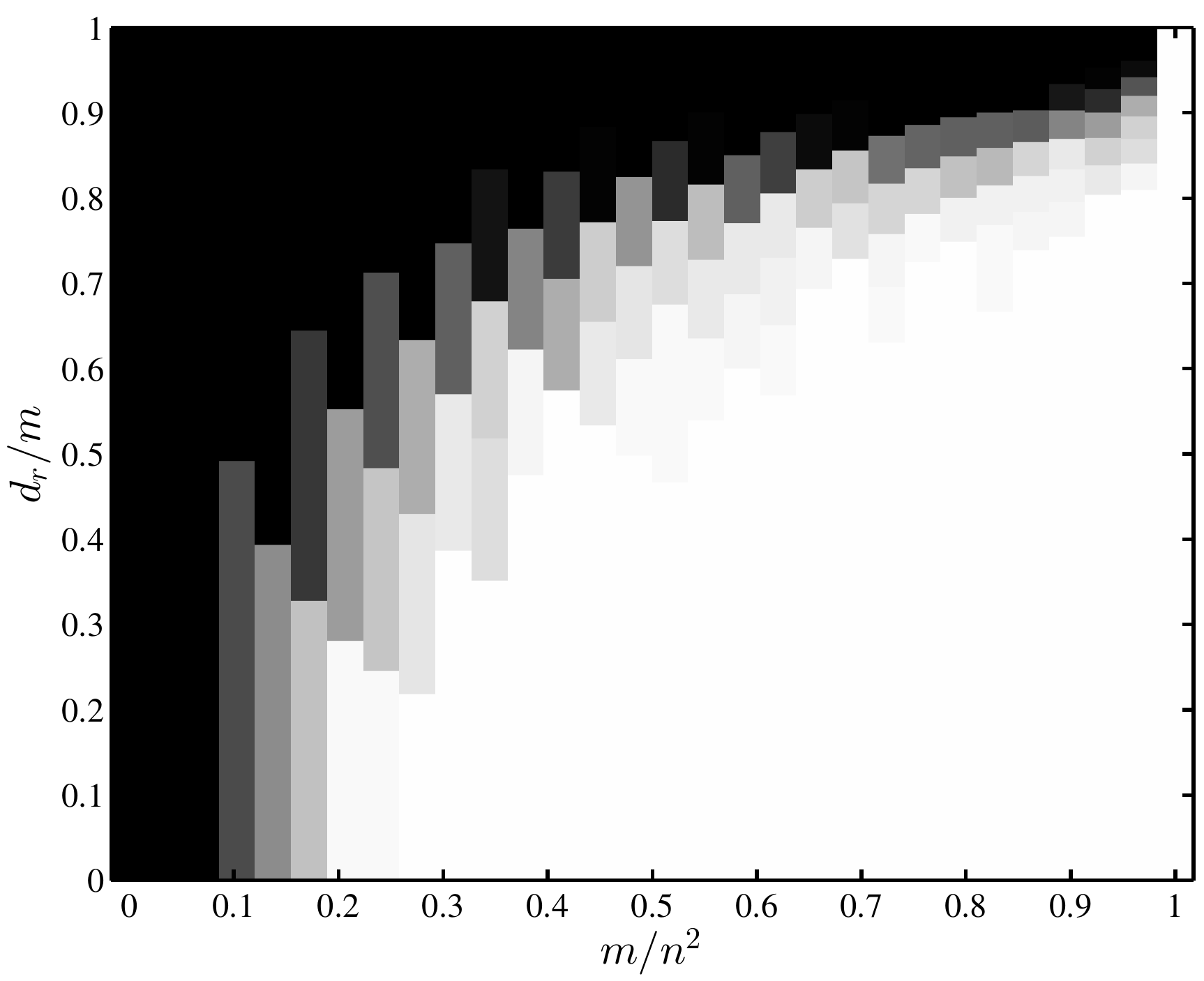}
        }%
        \subfigure[]{%
                \label{fig:DevMC4}
                \includegraphics[width=0.23\textwidth]{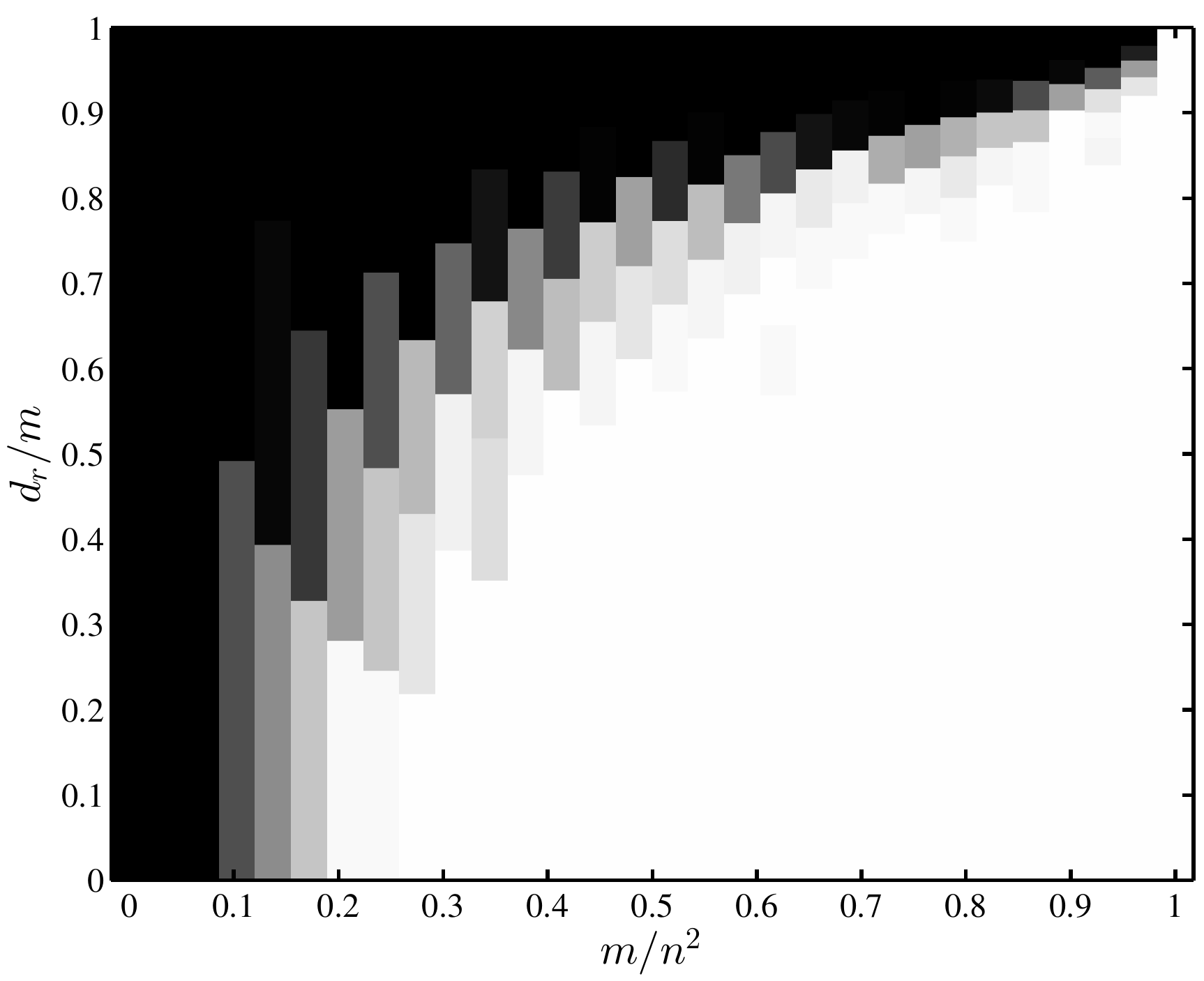}
        }%
        \vspace{-0.2cm}
        \caption{Phase transition plots for the ICRA algorithm in solving the MC problem as it proceeds with finer approximations of the rank function. (a) corresponds to the NNM which is used to initialize ICRA, and (b)-(d) correspond to the next three consecutive iterations of the external loop. Simulations are performed 50 times. Gray-scale color of each cell indicates the rate of perfect recovery. White denotes 100\% recovery rate, and black denotes 0\% recovery rate. A recovery is perfect if the $\RSNR$ is greater than 60 dB.} \label{fig:PhTransMC}
\end{figure}

\begin{figure}[tb]
        \subfigure[]{%
                \label{fig:DevRM1}
                \includegraphics[width=0.23\textwidth]{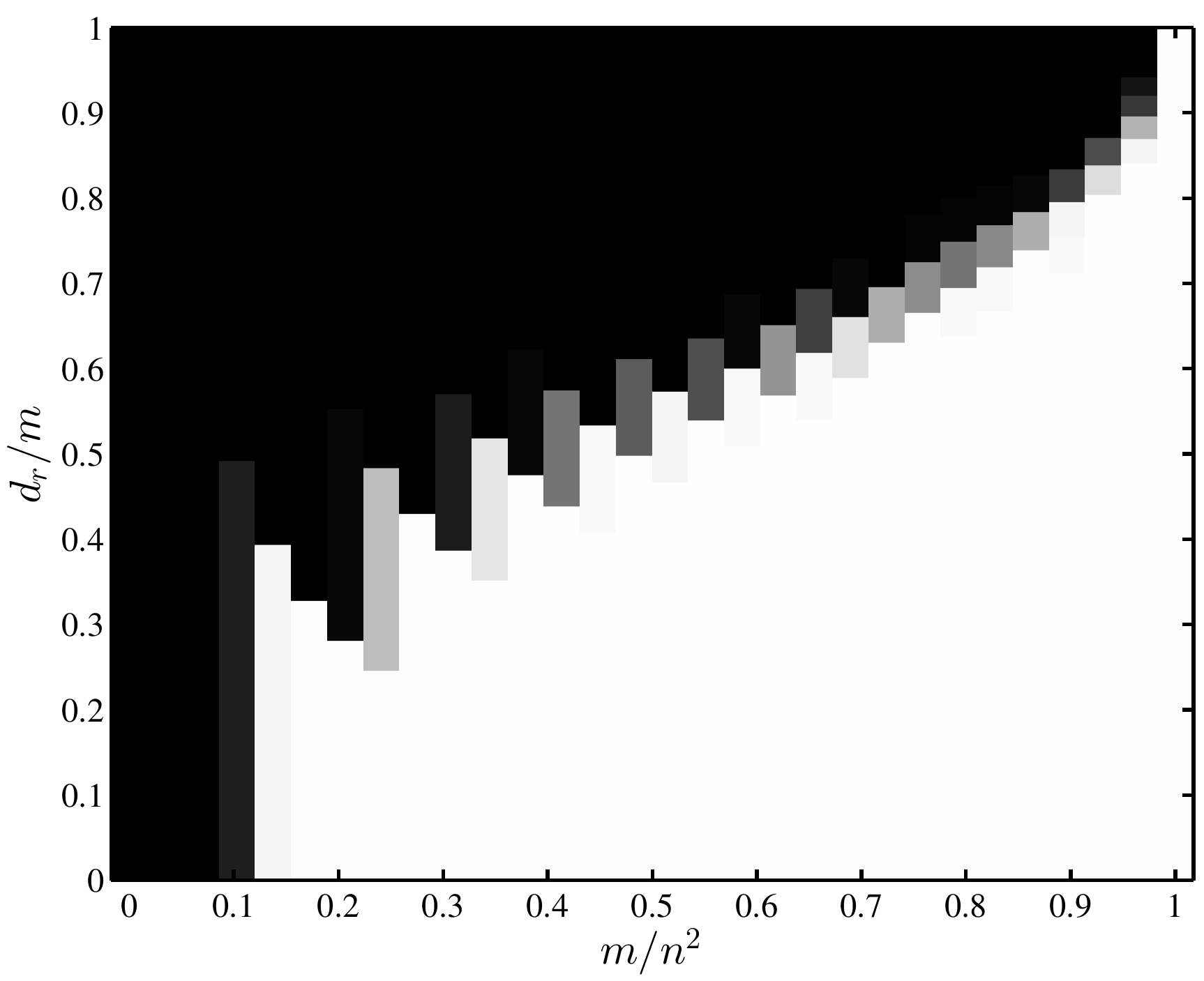}
        }%
        \subfigure[]{%
                \label{fig:DevRM2}
                \includegraphics[width=0.23\textwidth]{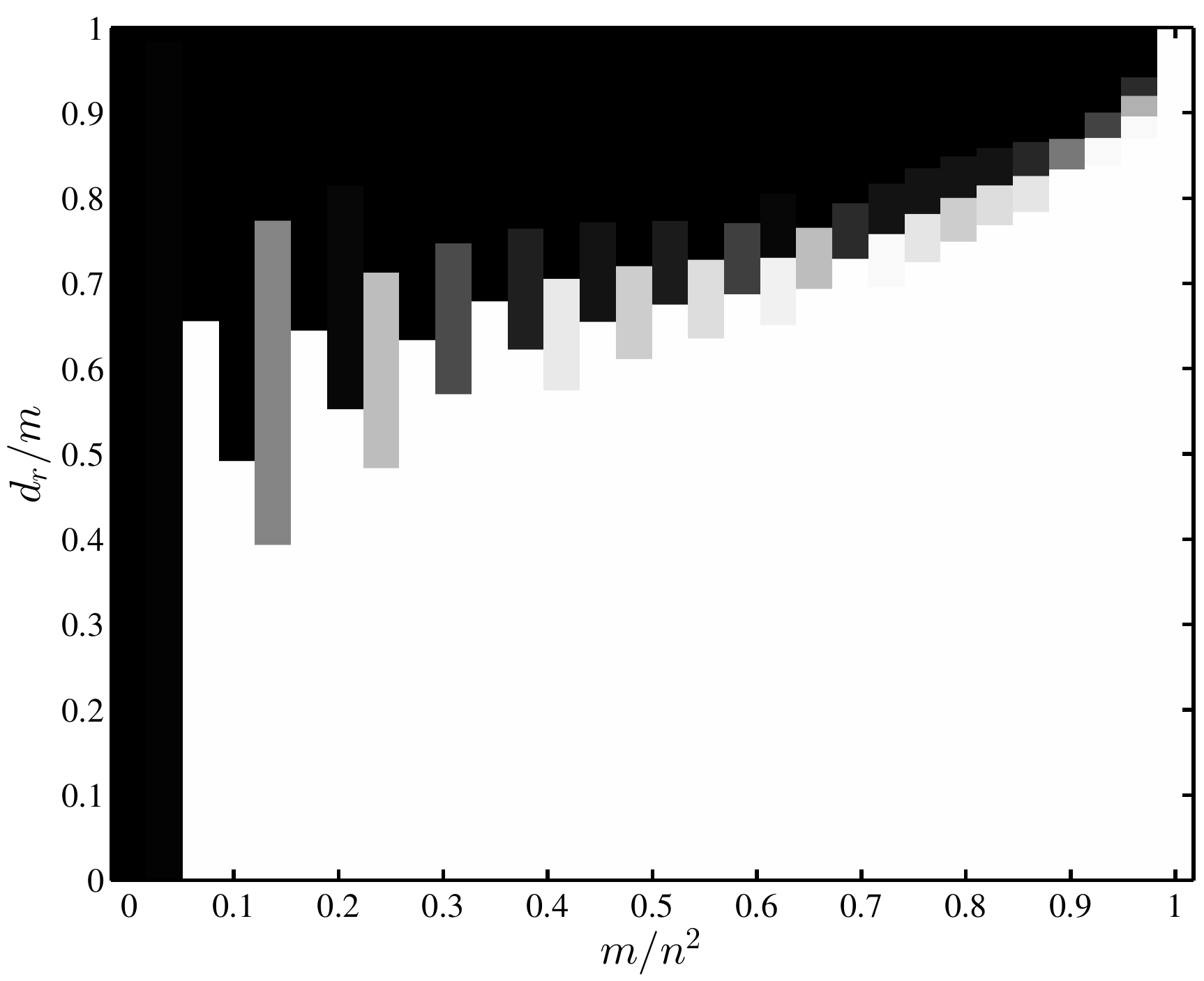}
        }%
        \vspace{-0.35cm}
        \subfigure[]{%
                \label{fig:DevRM3}
                \includegraphics[width=0.23\textwidth]{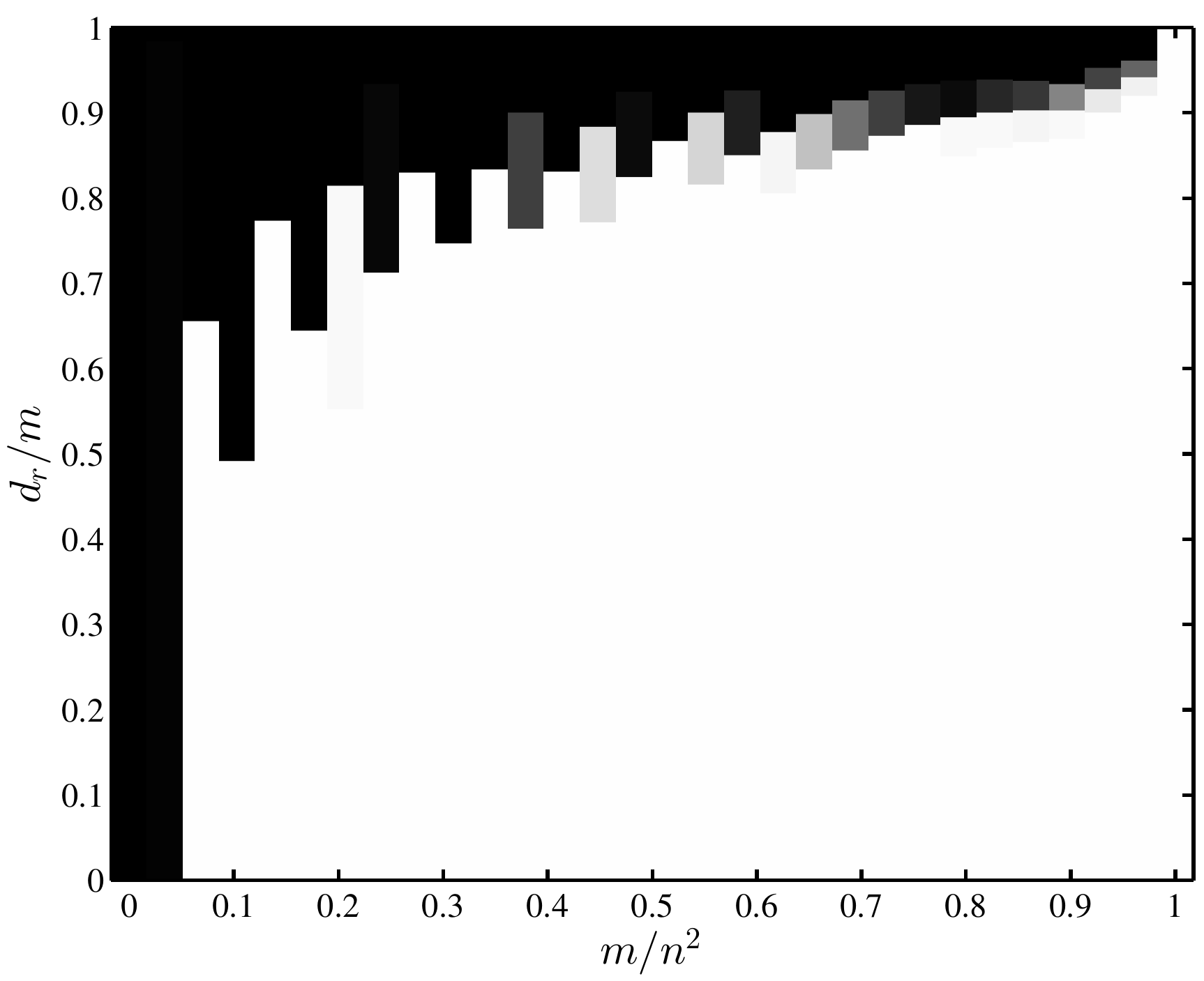}
        }%
        \subfigure[]{%
                \label{fig:DevRM4}
                \includegraphics[width=0.23\textwidth]{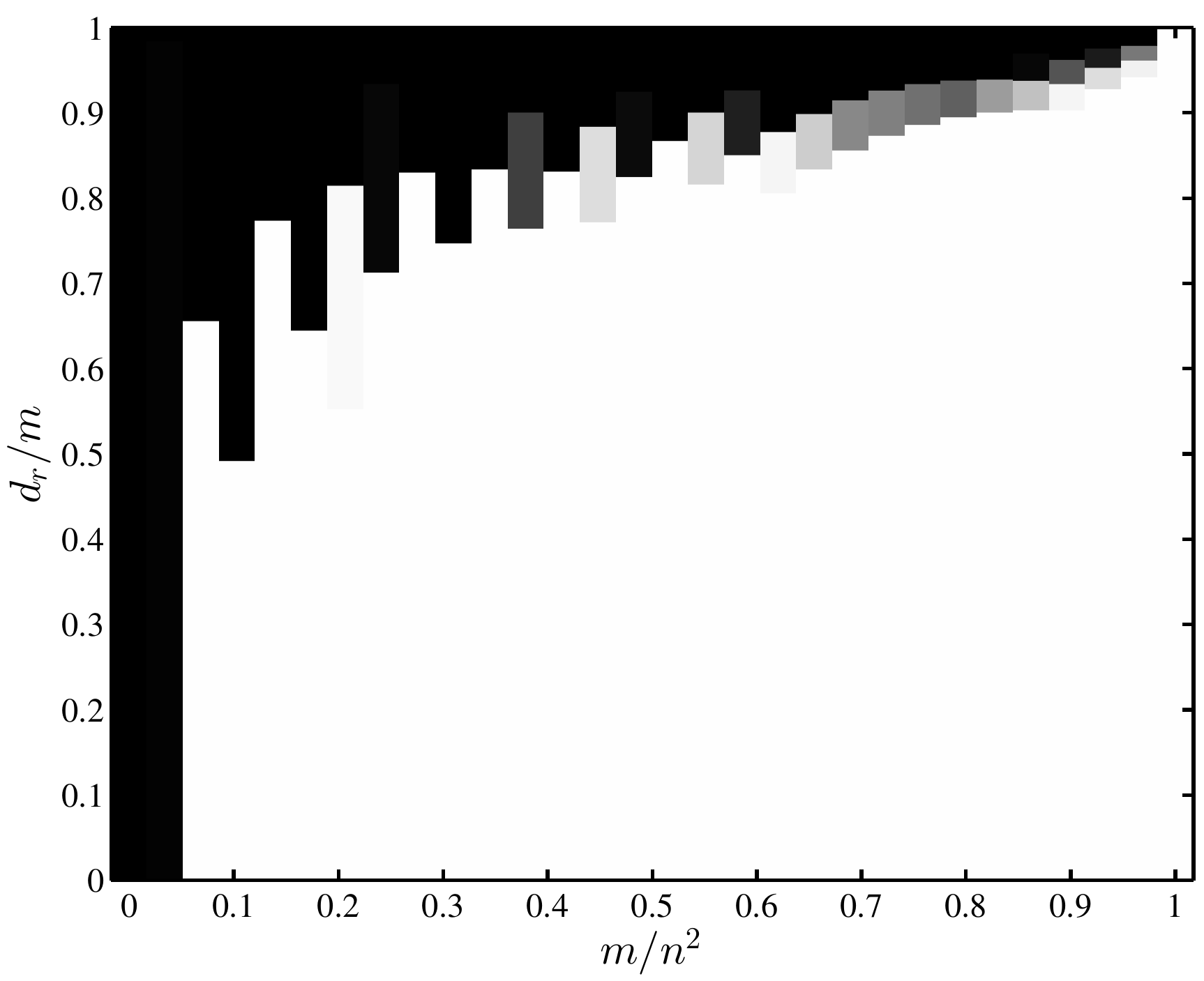}
        }%
        \vspace{-0.2cm}
        \caption{Phase transition plots for the ICRA algorithm in solving the ARM problem as it proceeds with finer approximations of the rank function. (a) corresponds to the NNM which is used to initialize ICRA, and (b)-(d) correspond to the next three consecutive iterations of the external loop. Other conditions are as in Figure \ref{fig:PhTransMC}.}\label{fig:PhTransRM}
\end{figure}

Figures \ref{fig:PhTransMC} and \ref{fig:PhTransRM} show the results of this experiment for ARM and MC problems. The gray color of each cell indicates the empirical recovery rate. White denotes perfect recovery in all trials, and black shows unsuccessful recovery for all trials. As clearly illustrated in these plots, when $\delta$ decreases the region of perfect recovery extends. Particularly, at two first iterations, the gain in the extension is more significant. Furthermore, our experiments shows that decreasing $\delta$ for more than four steps does not boost the performance meaningfully.

\begin{figure*}[tb!]
        \centering
        \subfigure[MC, $r = 2$.]{%
                \label{fig:MC1}
                \includegraphics[width=0.49\textwidth]{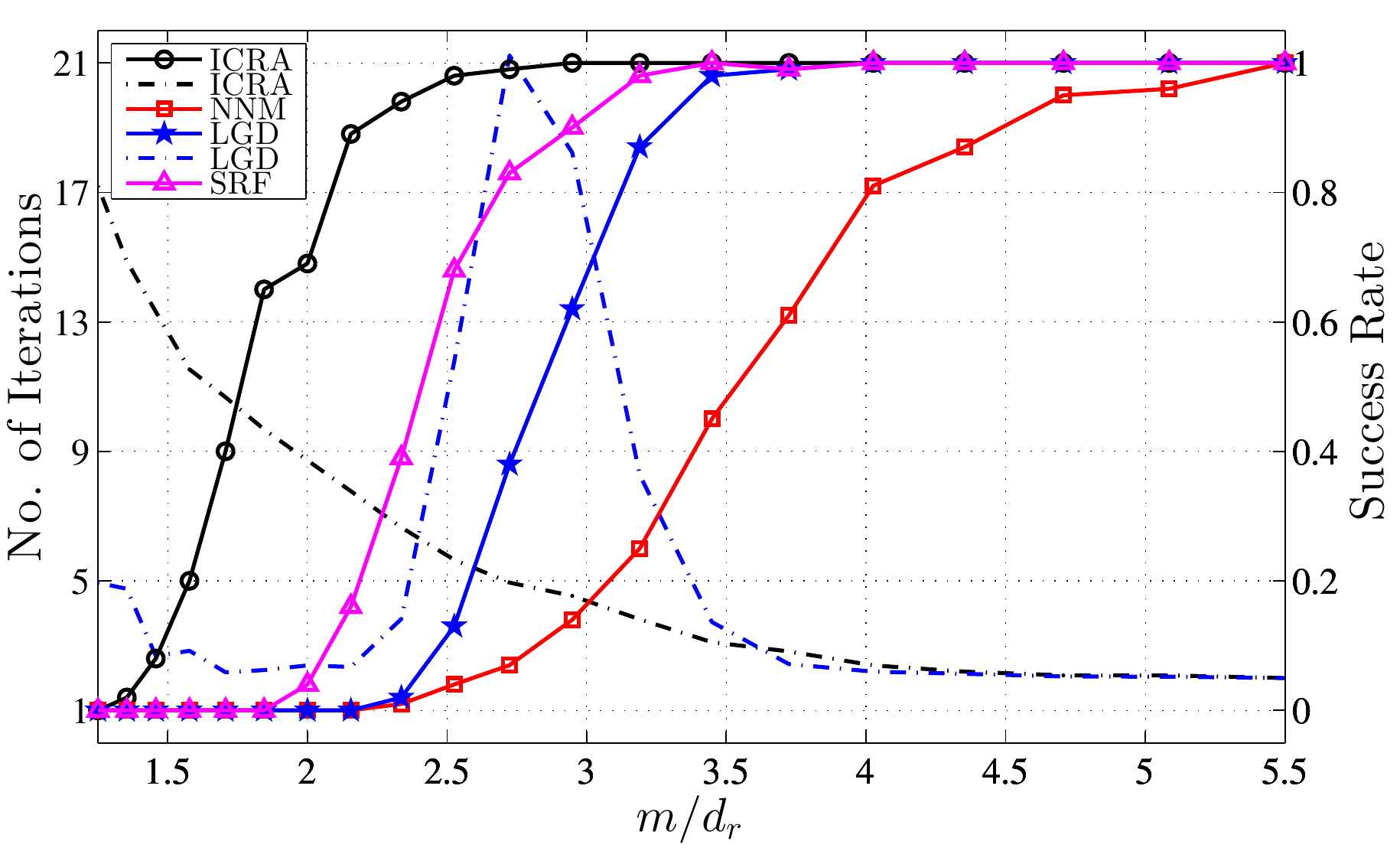}
        }%
        \subfigure[MC, $r = 5$.]{%
                \label{fig:MC2}
                \includegraphics[width=0.49\textwidth]{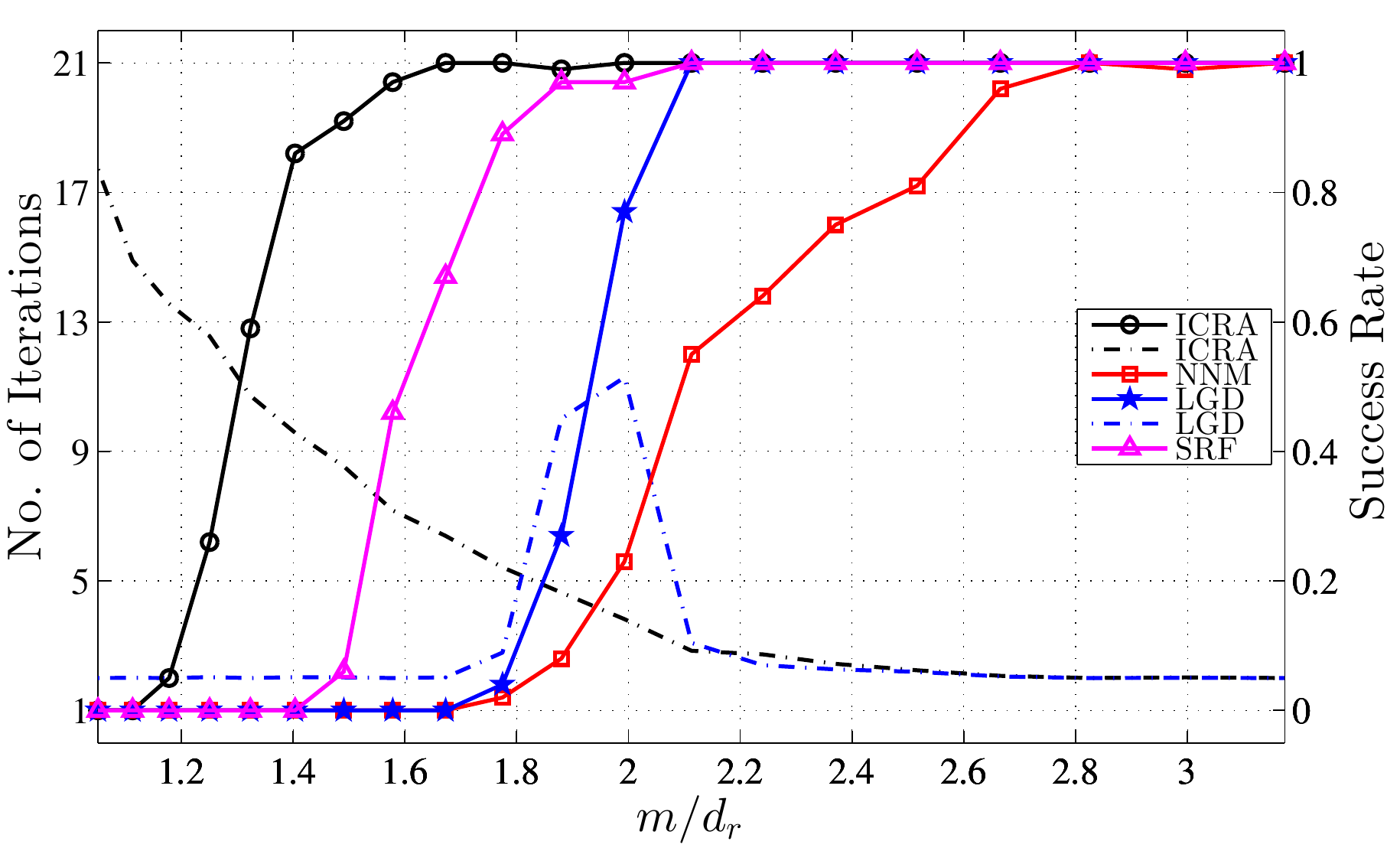}
        }%
        \vspace{-0.35cm}
        \subfigure[MC, $r = 10$.]{%
                \label{fig:MC3}
                \includegraphics[width=0.49\textwidth]{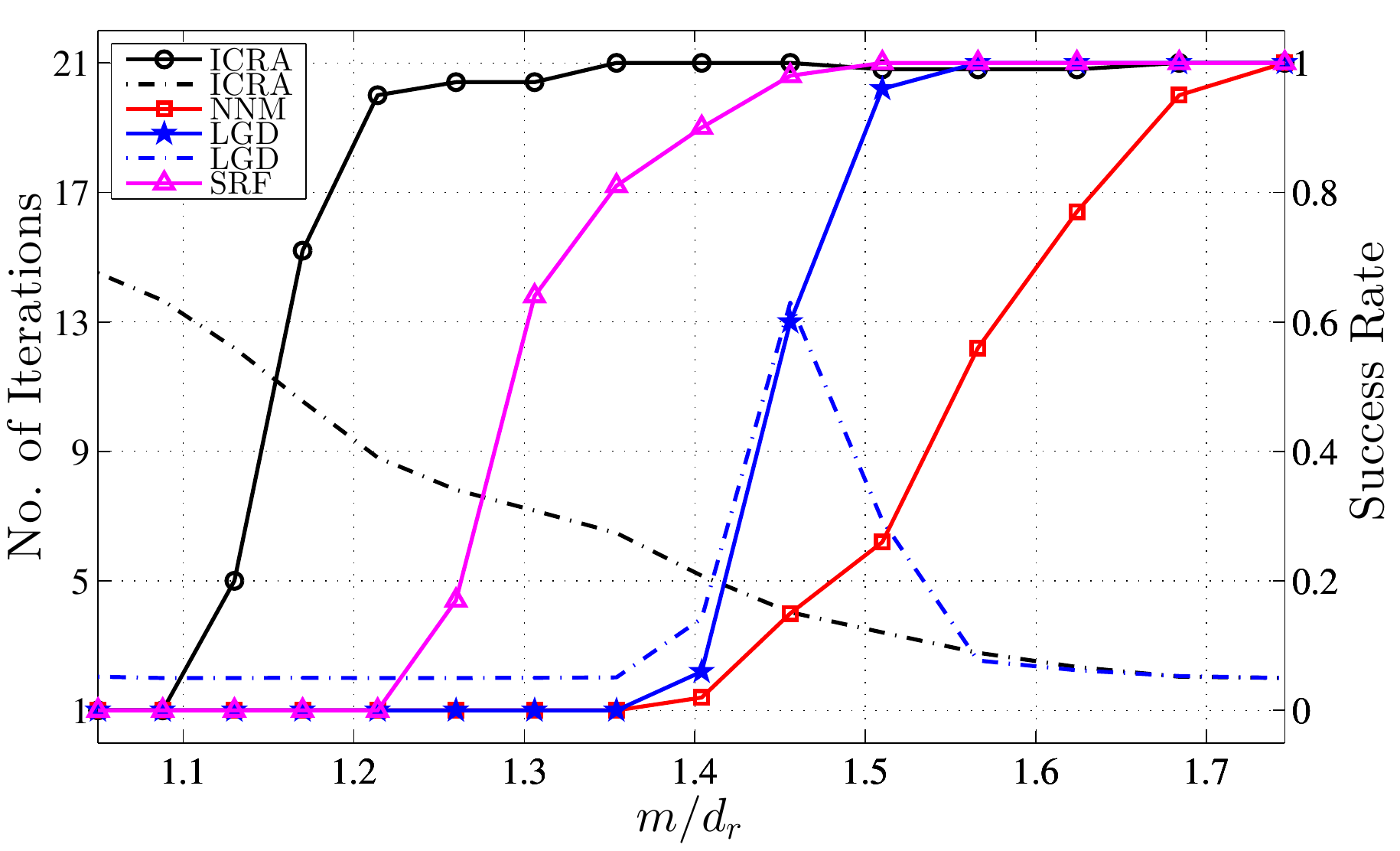}
        }%
        \subfigure[ARM, $r = 2$.]{%
                \label{fig:RM1}
                \includegraphics[width=0.49\textwidth]{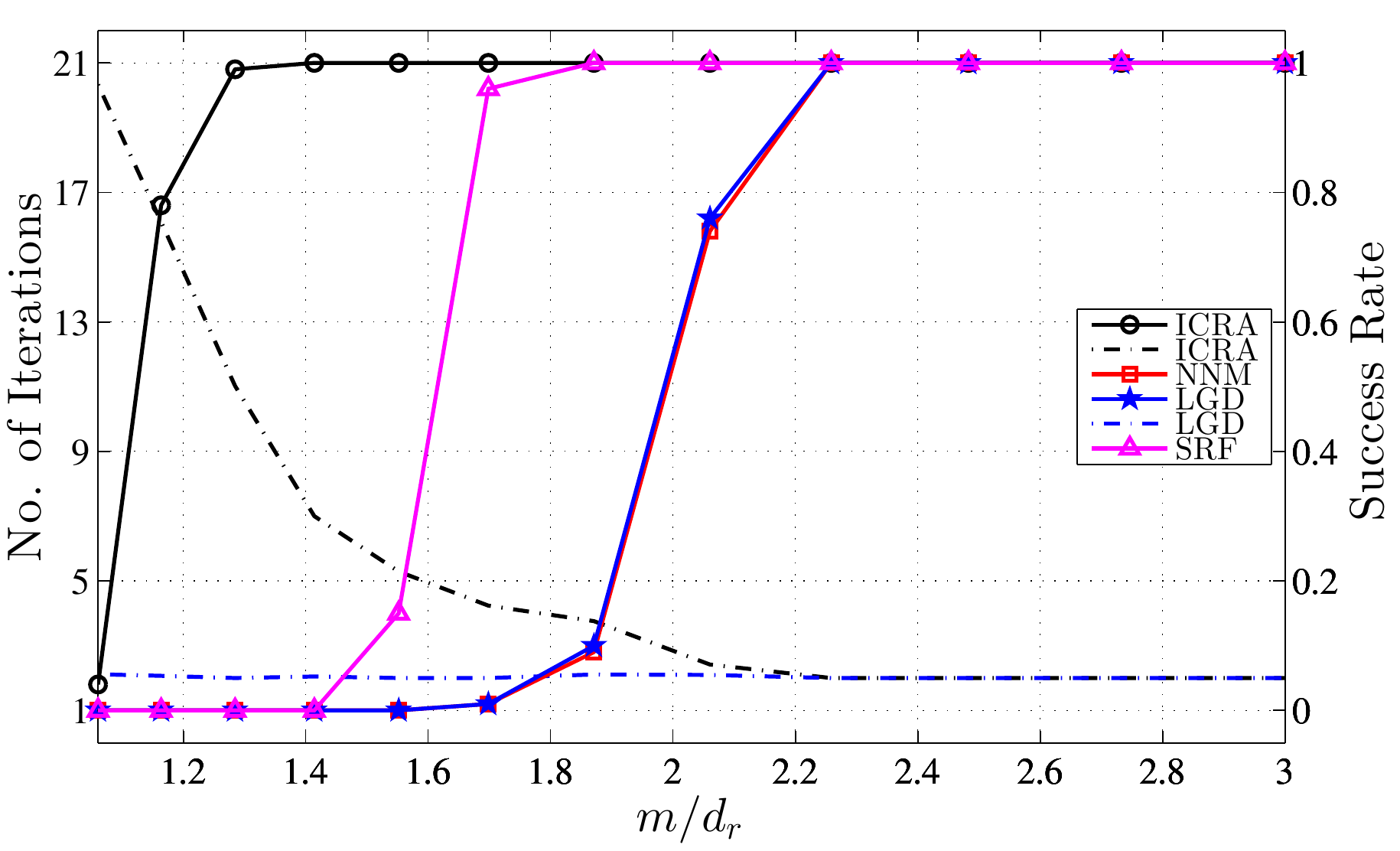}
        }%
        \vspace{-0.35cm}
        \subfigure[ARM, $r = 5$.]{%
                \label{fig:RM2}
                \includegraphics[width=0.49\textwidth]{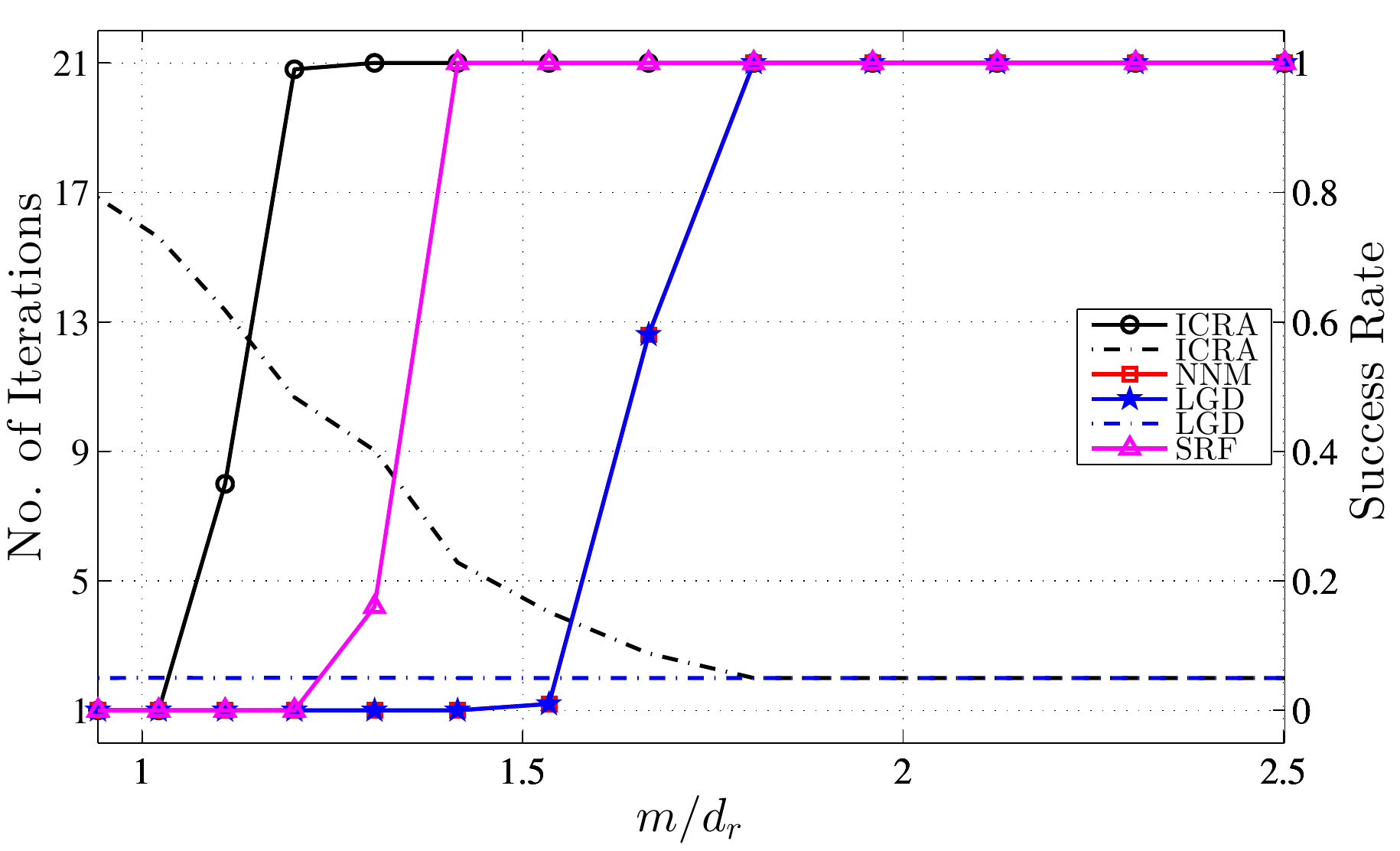}
        }%
        \subfigure[ARM, $r = 10$.]{%
                \label{fig:RM3}
                \includegraphics[width=0.49\textwidth]{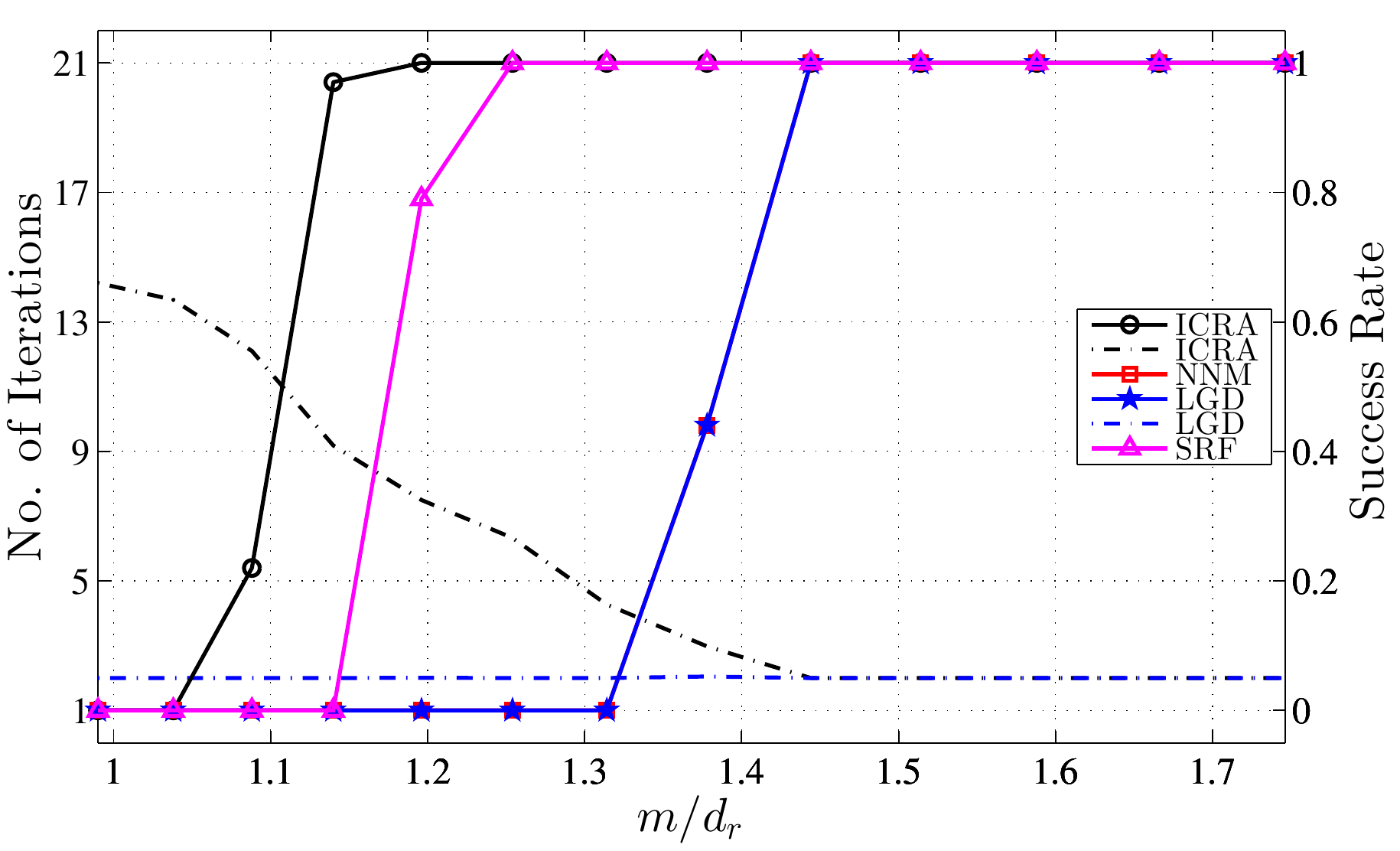}
        }%
        \vspace{-0.35cm}
        \subfigure[ARM, $r = 15$.]{%
                \label{fig:RM4}
                \includegraphics[width=0.49\textwidth]{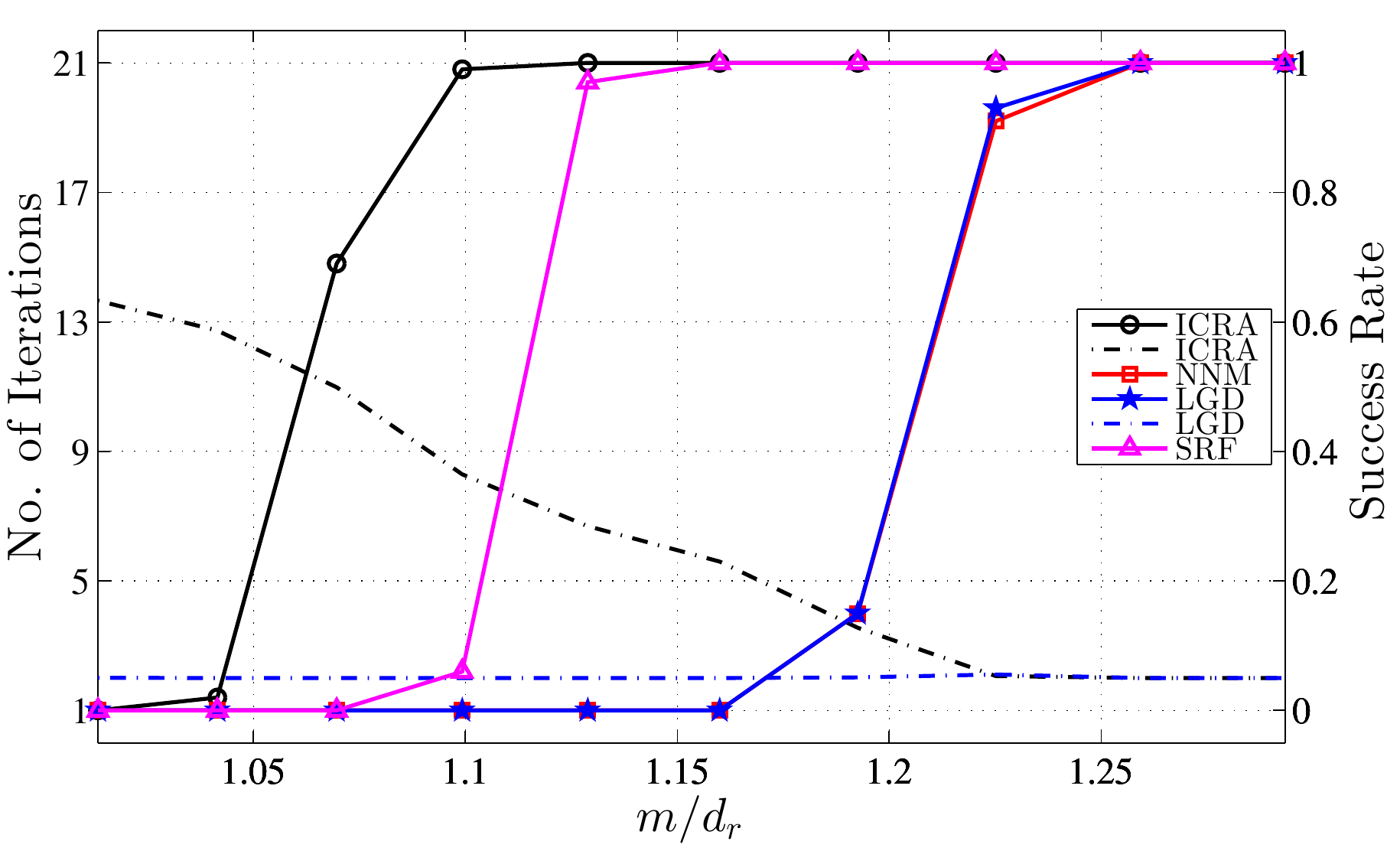}
        }%
        \subfigure[ARM, $r = 20$.]{%
                \label{fig:RM5}
                \includegraphics[width=0.49\textwidth]{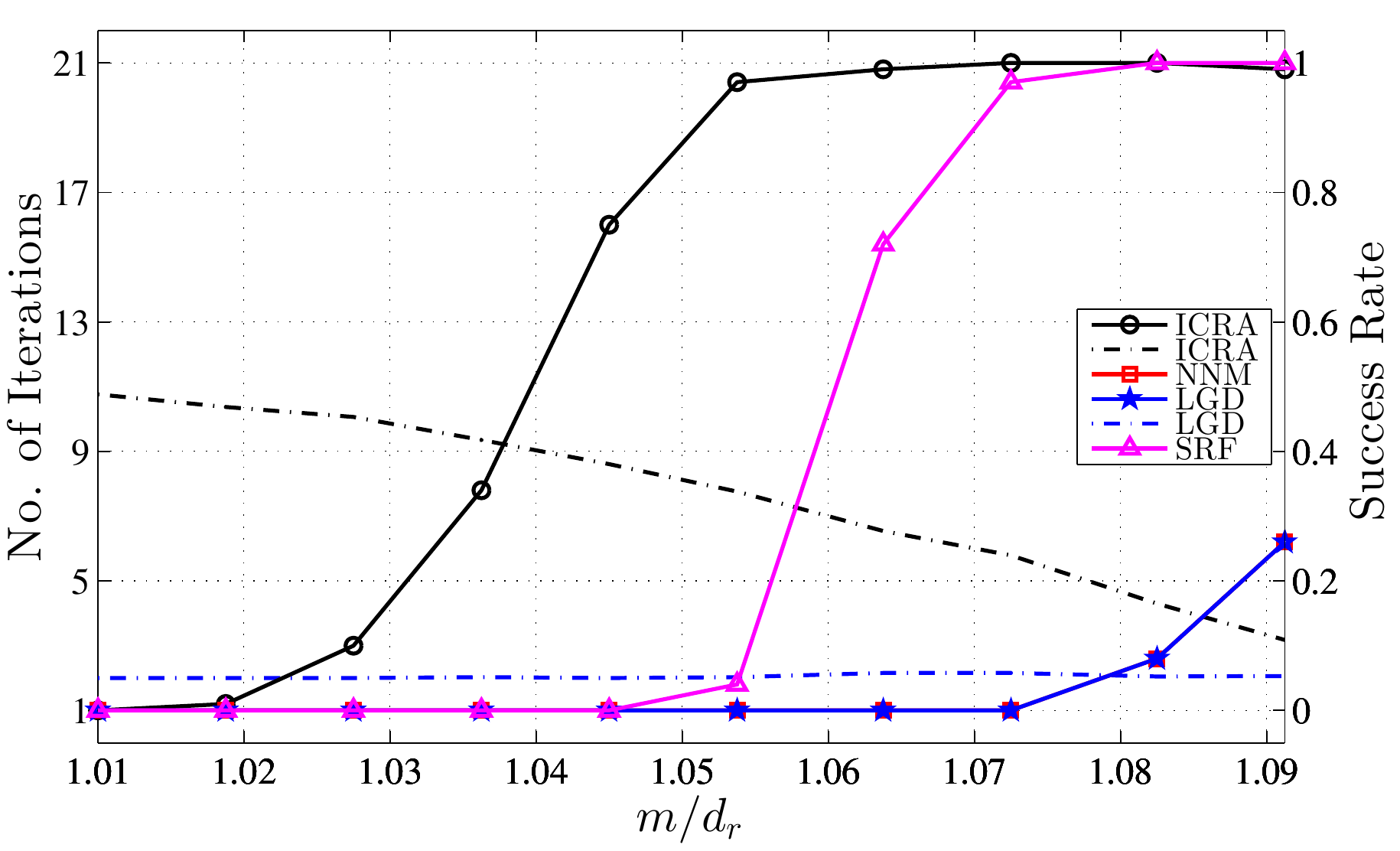}
        }%
        \vspace{-0.2cm}
        \caption{Comparison of the ICRA algorithm to the SRF \cite{MaleBAJ14}, LGD \cite{FazeHB03}, and NNM methods in terms of success rate and complexity. In these plots, the left-hand side vertical axis denotes the number of iterations each algorithm, except for SRF and NNM, needs to converge. In addition, the right-hand side vertical axis display the so-called success rate. A solid trace represents the success rate of an algorithm, and a dashed trace shows the number of SDPs the same algorithm used to find the solutions. Trials are repeated 100 times, and results are averaged over them.}\label{fig:AlgComp}
\end{figure*}

\emph{Experiment 3.} In this experiment, the ICRA algorithm is compared to NNM, LGD, and SRF methods in solving ARM and MC problems defined in \eqref{RM} and \eqref{MC}, respectively. Two criteria are used to this end: success rate and computational complexity. We declare an algorithm to be successful in recovery of the solution if $\SNR$ is greater than or equal to 60 dB. Consequently, the success rate of an algorithm denotes the number of times it successfully recovered the solution divided by the total number of trials, which is equal to 100 herein. Furthermore, the number of SDPs each algorithm, except SRF, needs to converge to a solution is reported as a measure of complexity. Although a rough estimate of complexity, this measure is independent of simulation hardware specifications and can give insight to the order of computational loads of the algorithms, as order of computation is fully understood for SDP solvers, see e.g. \cite{TohTT99}. We exclude SRF from this complexity comparison because it has an efficient implementation, whereas ICRA is realized by CVX as a proof-of-concept version. In addition, other competitors are also implementable by SDP, while SRF is not.

No stopping rule is specified in \cite{FazeHB03} for the LGD method, and we use the distance between two consecutive iterations to terminate it. To be precise, if $d = \| \Xbh_i - \Xbh_{i-1} \|_F / \|\Xbh_{i-1} \|_F  \leq \texttt{tol}$, where $\Xbh_i$ is the solution at the $i$th iteration, then the final solution is $\Xbh_i$. In all the comparisons, \texttt{tol} is set to $10^{-4}$ since we observed empirically that decreasing \texttt{tol} to smaller values only increases the number of LGD iterations, whereas $\SNR$ does not boost meaningfully. The SRF algorithm is executed with $c = 0.85, \mu = 1, L=8, \text{and } \epsilon = 10^{-5}$.

Figure \ref{fig:MC1}-\ref{fig:MC3} plots the success rate for ICRA, SRF, LGD, and NNM as well as number of SDP iterations for ICRA and LGD as a function of $m/d_r$ in solving MC problems with $r = 2,5, \text{ and } 10$, respectively. In these plots, the left-hand side vertical axis shows the average number of SDPs used to obtain the final solution, and the right-hand side vertical axis displays the success rate. Furthermore, a solid trace depicts the success rate of an algorithm, while the same color dashed trace shows the number of SDP iterations of the same algorithm. For instance, the black solid trace shows the success rate for the ICRA algorithm, and the dashed black one displays its total number of iterations. NNM method always gives a solution after execution of an SDP, so, to have more organized plots, this result is not shown.

It is clear from these results that, for the MC problems, ICRA can recover the solutions with considerably smaller number of measurements, and SRF stands in the second place of this comparison. Particularly, when $r$ equals to 10 with number of measurements less than 1.2 times of the matrix degrees of the freedom, solutions can be recovered by ICRA  with a recovery rate close to 1. So far as the complexity of ICRA is concerned, while average number of iterations can exceed 17, when $m$ increases toward values in which success rate is about 1, number of iterations continuously decreases and becomes equal to 2 when LGD starts to recover solutions with success rate of 100\%. Also, when LGD starts to recover the solutions, its number of iterations suddenly increases up to 21 for $r=2$, whereas 5 iterations in average suffice for ICRA to converge.

The strength of ICRA in ARM is also shown in Figure \ref{fig:RM1}-\ref{fig:RM5}. Under the same conditions as explained before, \eqref{RM} is solved for $r= 2,5,10,15,$ and $20$. To sum up the results, LGD and NNM have very close success rate in all simulations, and ICRA consistently outperforms both of them. As $r$ increases, the minimum $m / d_r$ in which ICRA  can perfectly recover solutions decreases and, in particular, it needs measurements just 5\% more than the solution degrees of freedom to recover with rate 1 when $r$ is equal to 20. Similar to the MC case, the average number of ICRA iterations is a declining function of $m$ and decreases to 2 when NNM and LGD starts to recover the solutions. In fact, since ICRA is initialized with the minimum nuclear-norm solution, when the global solution is attainable by nuclear norm minimization, ICRA maintains this solution and terminates after two iterations. This may be justified as follows. From Theorem \ref{Sup}, we expect that if \eqref{NNM} and \eqref{RM} share the same global solution, \eqref{Fmin} also share the same minimizer. Moreover, Theorem \ref{LConvThm} guarantees the convergence since ICRA is initialized by the global solution and the cost function does not increase at any iteration.

These experiments demonstrate that even though our performance analysis predicts that, in comparison to NNM, ICRA requires less or equal number of measurements to uniquely recover the solutions, strictly smaller number of measurements suffice for its success. Furthermore, it seems that the proposed approach for minimizing \eqref{FminPos} can find a global minimum in a wide range of $m$'s at the presented numerical examples.

\section{Conclusion} \label{sec:Con}
The problem of approximation of $\rank(\Xb)$ in ARM and MC settings was considered by formulating it as $\rank(\Xb) = \sumn u(\sigma_i(\Xb))$. To simplify this task, we focused on the approximation of the unit step function and proposed a class of subadditive functions which are closely match the unit step. The concavity and differentiability of the resulting matrix functions were characterized, proving that they are concave and differentiable for PSD matrices. Using a lemma from \cite{FazeHB03}, we generalized the concave approximation to arbitrary nonsquare matrices. To handle the nonconvexity of the optimization problem, we used a series of optimizations, where the quality of the approximation is successively increased. Furthermore, to theoretically support the proposed algorithm, we presented a theorem proving the superiority of the proposed approximation to NNM. Then we examined the performance of the ICRA algorithm via numerical examples in both ARM and MC problems. These examples showed that though the computational complexity is high in comparison to NNM, LGD, and SRF, ICRA can recover low-rank matrices with number of measurements close to the intrinsic unique representation lower-bound. The decrease in the number of measurements, in comparison to NNM, was up to 50\% in the performed numerical simulations.

\section*{Acknowledgment}
The authors wish to thank Dr. Arash Amini for many fruitful discussions and proof-reading of the manuscript. They also would like to thank the anonymous reviewers for their helpful comments.

\bibliography{SepSrc}
\bibliographystyle{IEEEbib}

\end{document}